\documentclass[a4paper,11pt,onecolumn,accepted=2026-09-06]{quantumarticle}
\pdfoutput=1
\usepackage[utf8]{inputenc}
\usepackage[T1]{fontenc}
\usepackage[english]{babel}
\usepackage{amsmath, amssymb, amsthm, mathrsfs, mathtools, bm}
\usepackage{graphicx}
\usepackage[numbers,sort&compress]{natbib}
\usepackage{physics}
\usepackage{subcaption}

\usepackage{xcolor}
\usepackage[bookmarksnumbered,colorlinks,citecolor=red,linkcolor=blue,hyperindex,linktocpage=true]{hyperref}
\usepackage[nameinlink,capitalise]{cleveref}

\newtheorem{theorem}{Theorem}
\newtheorem*{theorem*}{Theorem}
\newtheorem{proposition}{Proposition}
\newtheorem{lemma}{Lemma}
\newtheorem{definition}{Definition}
\newtheorem{assumption}{Assumption}
\newtheorem{remark}{Remark}

\newcommand{\Hil}{\mathcal{H}}
\newcommand{\Anc}{\mathcal{H}_{\mathcal{A}}}
\newcommand{\E}{\mathbb{E}}

\newcommand{\dt}{{\Delta t}}

\newcommand{\co}{{\mathcal{O} }}

\newcommand{\mat}[1]{\begin{pmatrix}#1 \\ \end{pmatrix}}

\title{Universal Dilation of Linear It\^o SDEs: Quantum Trajectories and Lindblad Simulation of Second Moments
}

\begin{document}

\author{Hsuan-Cheng Wu}
\email{wu.hsuancheng@psu.edu}
\author{Xiantao Li}
\email{xiantao.li@psu.edu}
\affiliation{Department of Mathematics, The Pennsylvania State University,
University Park, PA 16802, USA}
\date{}

\maketitle

\begin{abstract}
We present a universal framework for simulating $N$-dimensional linear It\^o stochastic differential equations (SDEs) on quantum computers with additive or multiplicative noises. Building on a unitary dilation technique, we establish a rigorous mapping from the general linear SDEs
\[
dX_t = A(t) X_t\,dt + \sum_{j=1}^J B_j(t)X_t\,dW_t^j
\]
to stochastic Schr\"odinger equations (SSE) on a dilated Hilbert space. Crucially, this embedding is pathwise exact in that the classical solution is recovered as a projection of the dilated quantum state for each fixed noise realization. 
We demonstrate that the resulting SSEs are {naturally implementable} on digital quantum processors, where the stochastic Wiener increments are encoded directly by preparing the ancillary qubits. Exploiting this physical mapping, we develop two algorithmic strategies: (1) a trajectory-based approach that uses sequential weak measurements to realize efficient stochastic integrators, including a second-order  scheme, and (2) an ensemble-based approach that maps moment evolution to a deterministic Lindblad quantum master equation, enabling simulation without Monte Carlo sampling. We provide error bounds based on a stochastic light-cone analysis and validate the framework with numerical experiments.
\end{abstract}

\section{Introduction}

Stochastic differential equations (SDEs) are ubiquitous models for dynamical systems subject to
fluctuating environments~\cite{oksendal2003stochastic}. Their applications span asset pricing in
quantitative finance~\cite{black1973pricing}, particle dynamics and turbulence in statistical
physics~\cite{pope2000turbulent,van1992stochastic}, continuous-time state estimation via Kalman
filters~\cite{kalman1960new,jazwinski1970stochastic}, and, more recently, generative modeling in
machine learning~\cite{song2020score}. In practice, the usefulness of these models relies on
efficient numerical integration~\cite{kloeden1992numerical}. As the system dimension $N$ increases,
classical methods frequently encounter the curse of dimensionality: the cost is amplified both by
the high-dimensional linear algebra and by the need to sample sufficiently many trajectories to
resolve statistics of interest. Related stochastic trajectory representations also arise in auxiliary-field quantum Monte Carlo: a Hubbard--Stratonovich transformation expresses the interacting imaginary-time propagator as an average over stochastic one-body evolutions in auxiliary fields, with fluctuating walker weights, providing a direct bridge between many-body ground-state calculations and linear stochastic dynamics~\cite{shi2021afqmc}.

Meanwhile, quantum algorithms have made significant progress on deterministic linear dynamics.
For linear ODEs and related evolution problems, one can leverage quantum linear systems methods
\cite{harrow2009quantum} and Hamiltonian simulation techniques
\cite{childs2021theory,childs2012hamiltonian,an2021time,gilyen2019quantum,berry2015simulating}, to design efficient simulation algorithms.
More generally, the Schr\"odingerisation paradigm maps linear ODE systems to time-dependent
Schr\"odinger equations, enabling the use of standard Hamiltonian-simulation primitives
\cite{jin2022quantum}. These developments motivate a parallel question: can one obtain an equally
native quantum representation for linear \emph{stochastic} dynamics?

A central difficulty is structural. The natural evolution of a quantum system is unitary, or, for
Markovian open systems \cite{breuer2002theory}, described by Lindblad master equations and their stochastic unravellings, also known as the 
stochastic Schr\"odinger equations, where the noise amplitude and dissipation are exactly balanced. General classical SDEs do not inherently satisfy these
constraints. In fact, the mismatch is already visible for linear It\^o systems: the drift matrix in
a classical SDE is typically non-Hermitian without possessing the specific dissipative form required
by an SSE.

\subsection{Problem setup: linear It\^o SDEs}

We consider an $N$-dimensional complex-valued process $X_t\in\mathbb C^N$ satisfying the linear It\^o
SDE
\begin{equation}\label{eq:SDE}
dX_t = \big(A(t)X_t + D(t)\big)\,dt + \sum_{j=1}^J \big(B_j(t)X_t + C_j(t)\big)\,dW_t^j,
\qquad X_0\in\mathbb C^N.
\end{equation}
Here $\{W_t^j\}_{j=1}^J$ are independent Wiener processes. We assume standard regularity conditions
(e.g.\ Lipschitz continuity and linear growth bounds) guaranteeing existence and uniqueness of a
strong solution with finite second moments~\cite{oksendal2003stochastic,kloeden1992numerical}.

Without loss of generality, it suffices to treat the homogeneous case (linear multiplicative noise),
\begin{equation}\label{eq:SDE0}
dX_t = A(t)X_t\,dt + \sum_{j=1}^J B_j(t)X_t\,dW_t^j,
\end{equation}
since additive terms can be embedded by augmenting the state with an auxiliary variable
$X_t^0\equiv 1$ and lifting \eqref{eq:SDE} to a homogeneous system in dimension $N+1$.

\subsection{ It\^o SDEs versus stochastic Schr\"odinger equations}

A natural quantum analogue of \eqref{eq:SDE0} is the stochastic Schr\"odinger equation (SSE) describing Markovian quantum
trajectories~\cite{breuer2002theory}:
\begin{equation}\label{eq:SSE-standard}
d\ket{\psi_t}
=
\Bigl(
  -iH(t)
  - \frac{1}{2}\sum_{j=1}^J V_j(t)^\dagger V_j(t)
\Bigr)\ket{\psi_t}\,dt
+ \sum_{j=1}^J V_j(t)\ket{\psi_t}\,dW_t^j,
\end{equation}
where $H(t)$ is Hermitian and $V_j(t)$ are coupling operators. The It\^o correction
$-\tfrac12\sum_j V_j^\dagger V_j$ is not optional: it enforces the characteristic open-system
structure, e.g.\ norm preservation in expectation for physical unravellings.

Comparing \eqref{eq:SDE0} and \eqref{eq:SSE-standard} reveals the obstruction. In general, the drift
$A(t)$ in a classical SDE \eqref{eq:SDE0} cannot be decomposed into $-iH(t)-\tfrac12\sum_j V_j(t)^\dagger V_j(t)$.
Equivalently, \cref{eq:SDE0} only has the same structure as \cref{eq:SSE-standard} if  the Hermitian matrix
\begin{equation}\label{eq:Kmat}
K(t) \;:=\; \frac{1}{2}\Big( A(t)+A(t)^\dagger + \sum_{j=1}^J B_j(t)^\dagger B_j(t)\Big)
\end{equation}
is zero. Consequently this term quantifies the failure of \eqref{eq:SDE0} to be compatible with a
standard SSE drift. In general it is neither generically small nor sign-definite, and therefore prevents a
direct identification of general linear SDEs with quantum stochastic models.

\subsection{A dilation viewpoint: compiling linear SDEs into open quantum dynamics}

In this work, we resolve the structural mismatch by treating \cref{eq:SDE0} as a \emph{template}
for linear stochastic dynamics rather than as a physical model. The key idea is a unitary
\emph{moment-matching dilation}: we embed the system into a larger Hilbert space
$\mathcal H_{\rm anc}\otimes \mathcal H_{\rm sys}$ and construct a dilated SSE whose coefficients are
chosen so that the original classical solution is recovered by a fixed projection of the dilated
trajectory, \emph{pathwise} for each noise realization.

This work should be viewed as a stochastic extension of the moment-matching dilation framework introduced in the earlier work \cite{li2025linear}. There, a general linear ODE was embedded into a unitary evolution on an enlarged Hilbert space, allowing the original non-unitary dynamics to be recovered through a fixed projection. The present paper extends this paradigm from deterministic linear systems to linear Itô SDEs by constructing a dilation into a stochastic Schrödinger equation (SSE), and consequently into a Lindblad evolution for second moments.

This dilation viewpoint has two immediate consequences that shape the rest of the paper.
First, it converts generic linear stochastic dynamics \cref{eq:SDE0} into native primitives of open quantum systems:
\emph{quantum trajectories}, which can be simulated via repeated interactions and measurements,  and
\emph{ensemble} evolution, governed by Lindblad dynamics for second moments. Second, it enables a finite-ancilla
implementation with provable control via a light-cone property, supporting long-time
simulation through segment-wise evolution and ancilla refresh, which has recently been constructed in \cite{li2025linear} for deterministic problems.

\subsection{Our Contribution: Quantum Simulation via Moment-Matching Dilation}

We resolve the structural mismatch between general linear It\^o SDEs \eqref{eq:SDE0} and physical
quantum evolutions  \eqref{eq:SSE-standard} by embedding  \eqref{eq:SDE0} into
a standard stochastic Schr\"odinger equation (SSE) on a dilated Hilbert space
\(\mathcal H_{\rm anc}\otimes\mathcal H_{\rm sys}\). Building on moment-matching dilation for
deterministic linear systems \cite{li2025linear}, we construct a dilated Hamiltonian and coupling operators so that the
classical solution is encoded in the dilated trajectory \(\ket{\psi_t(\omega)}\) satisfying an SSE system \eqref{eq:SSE-standard} and is recovered
\emph{pathwise} by a fixed linear readout:
\begin{equation}\label{eq:intro-pathwise-recovery}
X_t(\omega) \;=\; (\bra{l}\otimes I)\ket{\psi_t(\omega)}.
\end{equation}
This embedding enables two complementary quantum simulation routes, aimed at different output tasks.

\paragraph{Algorithm I: second-moment weak simulation. }
Many target quantities are quadratic, e.g.,
\(\mathbb E[X_T^\dagger O X_T]=\tr(O\Sigma_T)\) with \(\Sigma_T=\mathbb E[X_T X_T^\dagger]\).
The dilated second moment \(\rho_t=\mathbb E[\ket{\psi_t}\!\bra{\psi_t}]\) satisfies a
deterministic Lindblad master equation on \(\mathcal H_{\rm anc}\otimes\mathcal H_{\rm sys}\). As a quantum channel, the solution $\rho_t$ can be efficiently simulated using existing Lindblad simulation algorithms. 
Consequently, quadratic statistics can be estimated by simulating this Lindblad dynamics (without
sampling trajectories), followed by a single observable estimation on the final state. To reach long
times \(T\), we use segmentation of length \(\tau=\Theta(1/K_{\max})\) with $K_{\max}$ being the norm of the matrix in \cref{eq:Kmat}, together with ancilla refresh, 
via OAA on a window projector, and amplitude tracking through segment-wise growth factors.

\paragraph{Algorithm II: pathwise trajectory simulation.}
For applications requiring sample paths or expectations of general nonlinear functions, we directly simulate the dilated
SSE \eqref{eq:SSE-standard} as a repeated-interaction circuit. In each time step, we \emph{presample} a discrete
approximation of the Wiener increment, encode this choice into the ancilla state, and
apply a fixed interaction unitary. The output is a single (unnormalized) quantum state proportional
to \(X_T(\omega)\) for the chosen noise realization. As in the Lindblad route, long-time simulation
uses segmentation with ancilla refresh and the non-unitary trajectory scaling is tracked by estimating growth factors from each segment.

Throughout the paper we treat Lindblad simulation and trajectory simulation as black-box primitives. The quantities $C_{\mathcal L,T}$ and $C_{\mathrm{traj},T}$ denote the cost of implementing the corresponding dilated evolution under any chosen simulation framework. Consequently, our complexity estimates characterize the additional overhead introduced by the dilation, segmentation, refresh, and amplitude-tracking procedures, independently of the particular oracle or block-encoding model used to represent the operators $A(t)$ and $B_j(t)$. 

\begin{theorem*}[Informal complexity: Algorithm I (Lindblad/second moments)]
Let \(C_{\mathcal L,T}\) denote the cost of simulating the dilated Lindblad dynamics for total time
\(T\) (including segmentation and refresh), and let \(K_{\max}=\sup_t\|K(t)\|\) so that the number of
segments satisfies \(L=\Theta(TK_{\max})\). Let \(\Lambda_T:=\tr(\Sigma_T)\).

For segment \(m\), 
define the segment trace-growth factor
\begin{equation}\label{eq:gm-informal-lindblad}
g_m
\;:=\;
\frac{\tr(\Sigma_{t_{m+1}})}{\tr(\Sigma_{t_m})}, \quad \Gamma \;:=\;\sum_{m=0}^{L-1}\frac{1}{\sqrt{g_m}}.
\end{equation}
Then one can estimate \(\mu=\tr(\Sigma_T O)\) to additive error \(\varepsilon\) using total cost
scaling as
\[
\widetilde{\co}\!\left(
C_{\mathcal L,T} \Gamma\frac{\Lambda_T}{\varepsilon} 
\right),
\]
up to polylogarithmic factors and constant refresh overhead.
\end{theorem*}

\begin{theorem*}[Informal complexity: Algorithm II (trajectory generation)]
Let \(C_{{\rm traj},T}\) denote the cost of implementing the presampled weak-integrator trajectory
circuit up to time \(T\), with \(L=\Theta(TK_{\max})\) segments of length \(\tau=\Theta(1/K_{\max})\).
Let \(\mathcal M_m^{(\tau)}\) denote the (random, presampled) linear segment map on \([t_m,t_{m+1}]\)
acting on the system state.

Define the segment growth factor as the ratio of squared amplitudes,
\begin{equation}\label{eq:gm-informal-traj}
g_m
\;:=\;
\frac{\|\mathcal M_m^{(\tau)}\ket{\psi_{t_m}}\|^2}{\|\ket{\psi_{t_m}}\|^2}, \quad 
\Gamma_1:=\sum_{m=0}^{L-1}\frac{1}{\sqrt{g_m}},
\quad
\Gamma_2:=\sum_{m=0}^{L-1}\frac{1}{g_m}.
\end{equation}
Then the algorithm outputs a single trajectory state proportional to \(X_T(\omega)\) for a presampled noise
realization \(\omega\), together with  estimates of its amplitude, with overall cost scaling as
\[
\widetilde{\mathcal O}\!\left(
C_{{\rm traj},T}\Big[\Gamma_1+\frac{L}{\varepsilon}\Big]
C_{{\rm traj},T}\Big[\Gamma_1+\frac{L}{\varepsilon}\Big]
\right).
\]
\end{theorem*}

\medskip

\subsection{Related works.}

\noindent \textbf{Deterministic linear dynamics.}
Quantum algorithms for deterministic linear ODE/PDE systems are by now well developed, typically
reducing time propagation to block-encodings and Hamiltonian simulation primitives (via LCU/QSVT),
or to quantum linear-systems subroutines in time-discretized formulations.  Representative examples
include  \cite{Ber14,BCOW17,BC22,childs2020quantum,krovi2023improved} and more recent refinements that improve precision dependence and broaden
the class of implementable (generally non-unitary) linear maps.  Closest in spirit to our dilation
viewpoint is \emph{Schr\"odingerisation}, which maps general linear evolution to a time-dependent
Schr\"odinger equation on a larger Hilbert space \cite{jin2022quantum,jin2023schrodingerisation_pra},
as well as linear combination of Hamiltonian simulation \cite{ALL23,ACL23} and {moment-matching} dilations for non-unitary linear dynamics
\cite{li2025linear}.  These techniques motivate the present work: our goal is to extend such
dilations from deterministic linear dynamics to \emph{stochastic} linear It\^o systems while
retaining a physically standard quantum-mechanical form.

\noindent \textbf{Quantum algorithms for SDEs via time discretization and PDE reformulations.}
One line of work treats SDE simulation by first discretizing time (e.g.\ Euler--Maruyama or higher
weak schemes) and then reducing the resulting random time-stepping to a deterministic quantum
evolution after the Brownian increments are \emph{presampled} 
\cite{jin2025quantum}.  In contrast, our formulation is intrinsically continuous-time: we embed
the SDE into a SSEs, i.e.\ a quantum-trajectory model, making explicit connections to the theory of open quantum systems. 

A different route replaces the SDE by a deterministic PDE for a probability density, i.e., the Fokker--Planck (Kolmogorov forward) equation.  This approach enables the use of quantum PDE solvers \cite{jin2024quantum}, but
the resulting complexity typically involves the {PDE discretization}, and hence
can inherit polynomial dependence on the SDE dimension \(N\) in generic settings.

\noindent\textbf{Complexity-theoretic perspective and nonlinear/noisy dynamics.}
Beyond algorithmic constructions, recent work indicates that SDE simulation captures the full power
of quantum computation in a precise complexity-theoretic sense.  Bravyi \emph{et al.} study quantum
simulation of noisy classical nonlinear dynamics and establish BQP-completeness for the SDEs simulation tasks. They 
propose a bosonic-operator encoding to achieve favorable dimension dependence in
structured regimes \cite{bravyi2025quantum}.

\noindent\textbf{Relation to quantum trajectories and unravellings.}
Our trajectory algorithm also connects to the longstanding quantum-jump/quantum-trajectory
literature, where Lindblad evolution is unraveled into stochastic pure-state evolutions (SSEs)
implemented by repeated interactions and measurements; see, e.g., the review \cite{plenio1998quantumjump}
and foundational developments in wave-function Monte Carlo methods.  The key distinction is that we
use the SSE {as a computational representation} of a {classical} linear SDE via our
moment-matching dilation, thereby turning generic linear stochastic dynamics into a standard open quantum system model that is amenable to modern Lindblad simulation and trajectory-generation algorithms.

\noindent\textbf{Quantum algorithms for the Lyapunov equation.}
The second moment of the SDEs \eqref{eq:SDE0} satisfies a Lyapunov equation for the covariance $\Sigma(t)$.
Benedetti \emph{et al.}~\cite{benedetti2025probabilistic} propose a probabilistic quantum algorithm to prepare a mixed state proportional to the steady-state solution of the Lyapunov equation.
In contrast, our approach embeds the \emph{time-dependent} second-moment dynamics induced by the SDE into a Lindblad evolution via a moment-matching dilation, enabling transient covariance estimation as well.

\bigskip

The remainder of the paper is organized as follows. In \cref{sec: dsde}, we introduce the preliminaries for moment-matching dilation and the setup for mapping to SSEs. In \cref{sec: fddil}, we present a specific dilation using a finite-dimensional tight-binding model and prove the finite-time accuracy.  In \cref{sec:lindblad,sec: qtraj}, we elaborate on the implementations of the algorithms on digital quantum devices. Numerical results that validate the error estimates are provided in \cref{sec: numsim}.

\section{Dilation of Stochastic Differential Equations}\label{sec: dsde}

\subsection{Preliminaries: moment-matching dilation for ODEs}

We recall the deterministic dilation from our earlier work \cite{li2025linear} on non-unitary linear ODEs. Here, we only state what we need for the stochastic extension. Consider the linear ODE on $\mathbb{C}^N$
\[
\dot{\bm x}(t) = L(t) \bm x(t),\qquad L\in\mathbb{C}^{N\times N}.
\]
$L$ can be uniquely decomposed as $L(t) = -iH(t) + K(t)$ whereas both $H$ and $K$ are Hermitian.

\begin{definition}[Moment-matching dilation] 
Let $\Anc$ be a complex ancillary Hilbert space. A triple $(F,\ket{r},\bra{l})$ with
\[
F:\Anc\to\Anc,\qquad F^\dagger=-F,
\]
is called a \emph{moment-matching dilation of order $\infty$} if
\begin{equation}\label{eq:mm-condition}
\bra{l} F^k \ket{r} = 1,\qquad \forall k\in\mathbb{N}_0.
\end{equation}
\end{definition}
To clarify the notation, let $\Hil$ be the original Hilbert space, let $I$ be the identity matrix on $\Hil$, and let $I_\mathcal{A}$ denote the identity on the ancilla Hilbert space $\Anc$. Specific examples of moment-fulfilling families can be found in \cite{li2025linear}.

\begin{theorem}[Deterministic moment-matching dilation]
Let $L=-iH+K$ and let $(F,\ket{r},\bra{l})$ be a moment-matching triple. Define the dilated Hamiltonian on $\Anc\otimes\Hil$ by
\begin{equation}\label{eq:H-dil-ODE}
\widetilde{H} := I_{\mathcal{A}} \otimes H + i\, F \otimes K.
\end{equation}
Then, for all $t\ge0$,
\begin{equation}\label{eq:ODE-dilation-identity}
\mathcal T e^{\int_0^t L(t') dt'} = (\bra{l}\otimes I)
\mathcal T 
e^{-i \int_0^t \widetilde{H}(t') dt'}\,(\ket{r}\otimes I).
\end{equation}
\end{theorem}

The proof is purely algebraic: expand $\mathcal T 
e^{-i \int_0^t \widetilde{H}(t') dt'}$ as Dyson series in $t$, use that each factor $I\otimes H + iF\otimes K$ is a polynomial in $F$, and then use \eqref{eq:mm-condition} to replace each $k$-fold $F$ and $I_\mathcal{A}$ by the scalar $1$ in the matrix element $(\bra{l}\otimes I)(\cdot)(\ket{r}\otimes I)$, thus recovering the Dyson series of $\mathcal T e^{\int_0^t L(t') dt'}$ term by term.

\subsection{The exact-mapping theorem for SDEs \texorpdfstring{\eqref{eq:SDE0}}{}}

We now extend the dilation technique for simulating ODEs to the simulation of SDEs.  To guarantee the existence of unique strong solutions and the validity of the higher-order stochastic expansions used in our derivation, we impose the following regularity conditions.

\begin{assumption}[Regularity of Coefficients]\label{ass:regularity}
The drift operator $A(t)$ and noise operators $B_j(t)$ are uniformly bounded and continuously differentiable functions of time on the interval $t \in [0, T]$. That is, there exists a constant $C > 0$ such that for all $t$ and $j$:
\[
\|A(t)\| + \|B_j(t)\| \le C \quad \text{and} \quad \|\dot{A}(t)\| + \|\dot{B}_j(t)\| \le C.
\]
\end{assumption}

Under \cref{ass:regularity}, the linear SDE~\eqref{eq:SDE0} satisfies the standard global Lipschitz and linear growth conditions. This ensures the existence of a unique strong solution $X_t$ adapted to the filtration $\mathcal{F}_t$, satisfying $\sup_{t\in[0,T]} \mathbb{E}\|X_t\|^2 < \infty$ \cite[Thm.~4.5.3]{kloeden1992numerical}. Furthermore, the $C^1$-regularity ensures that the stochastic Taylor expansion converges in the mean-square sense.

We summarize the stability properties of the exact solution below \cite{khasminskii2011stochastic}.

\begin{proposition}[Mean-Square Stability and Growth]\label{prop:stability}
Let $X_t$ be the solution to \eqref{eq:SDE0} under \cref{ass:regularity}. Define the Hermitian Lyapunov matrix:
\begin{equation}\label{eq:Kmat'}
    K(t) := \frac{1}{2}\left( A(t) + A(t)^\dagger + \sum_{j=1}^J B_j(t)^\dagger B_j(t) \right).
\end{equation}
Then, the second moment evolves according to the differential equation
\begin{equation}\label{eq:moment-ode}
    \frac{d}{dt}\mathbb{E}\left[ \|X_t\|^2 \right] = 2 \mathbb{E}\left[\langle X_t, K(t)\,X_t\rangle\right].
\end{equation}
Consequently, the growth of the system is strictly controlled by the maximal eigenvalue of $K(t)$. If $K(t) \preceq \gamma(t) I$ for a scalar function $\gamma(t)$, we have the a priori bound:
\begin{equation}\label{eq:growth-bound}
    \mathbb{E}\left[\|X_t\|^2\right] \le \exp\left(2\int_0^t \gamma(s)\,ds\right)\,\mathbb{E}\left[\|X_0\|^2\right].
\end{equation}
In particular, if $K(t)$ is uniformly negative definite, the system is exponentially mean-square stable.
\end{proposition}

\bigskip 

We now elaborate on the dilation procedure. Starting from the SDE \eqref{eq:SDE0}, we fix an ancillary space $\Anc$ and a moment-matching triple $(F,\ket{r},\bra{l})$ as above. We start by defining,
\begin{equation}
    L(t):= A(t) + \frac{1}{2}\sum_{j=1}^J B_j^\dagger(t) B_j(t)
\end{equation}
from the drift and noise coefficients in the SDE system \eqref{eq:SDE0}. 

Next we split the operator $L(t)$ by defining the Hermitian matrix $H$ and $K$ by, 
\begin{equation}\label{eq:residual-drift}
L(t):= -iH(t) + K(t).
\end{equation}
One can verify that the hermitian part here is the same as that in \cref{eq:Kmat}.
Importantly, when $L(t) $ is skew-Hermitian, or equivalently $K(t)\equiv 0$, the original SDE system \eqref{eq:SDE0} coincides with the stochastic Schr\"odinger equation in \cref{eq:linSSE}. To be able to simulate the case when $K(t)\neq 0$,  we extend the dilation method, by extending the operators in dilated Hilbert space as follows,
\begin{equation}\label{H-dil}
    \widetilde{H} \coloneqq I_{\mathcal{A}} \otimes H(t) + iF \otimes K(t).
\end{equation}
We now extend the operators $B_j$ by a direct dilation,
\begin{equation}
    V_{j}(t) \coloneqq I_{\mathcal{A}} \otimes B_{j}(t),\quad j=1,\cdots,J.
\end{equation}
For the convenience of the presentation, we also define,
\begin{equation}
    V_{0}(t) \coloneqq -i\widetilde{H} - \frac{1}{2}\sum_{j=1}^{J} V_j^{\dagger} V_j,
\end{equation}
which will become the non-Hermitian part of the SSE \eqref{eq:SSE-standard}.

We will show that under the moment conditions \eqref{eq:mm-condition}, the dilation of these operators yields an SSE of the form,
\begin{equation}\label{eq:dilated-SSE}
d\ket{\psi_t}
=
V_0(t)\ket{\psi_t}\,dt
+ \sum_{j=1}^J {V_j}(t)\ket{\psi_t}\,dW_t^j,
\end{equation}
which has the same structure as \eqref{eq:SSE-standard}.

We set the initial condition of the dilated SSE \eqref{eq:dilated-SSE} to
\begin{equation}\label{Psi-init-val}
    \ket{\psi_0} = \ket{r}\otimes X_0,
\end{equation}
which can be easily prepared as a factored state. 

\begin{theorem}[Exact recovery of the linear SDE]\label{thm:main}
Let $X_t$ be the unique strong solution of the linear SDE \eqref{eq:SDE0} and $\ket{\psi_t}$ be the strong solution of the dilated SSE \eqref{eq:dilated-SSE} with initial data \eqref{Psi-init-val}. Under the assumptions on the coefficient matrices $A(t)$ and $B_j(t)$ in \cref{ass:regularity}, and that $(F,\ket{r},\bra{l})$ is a moment-matching triple according to \cref{eq:mm-condition},  for every $t\ge0$ the following identity holds almost surely:
\begin{equation}\label{eq:recovery}
X_t = (\bra{l}\otimes I)\ket{\psi_t}.
\end{equation}
Equivalently, for each fixed sample path $\omega\in\Omega$, the projected process $t\mapsto (\bra{l}\otimes I)\psi_t(\omega)$ is the exact solution of \eqref{eq:SDE0} with the same Wiener trajectory.
\end{theorem}
The proof is presented in \cref{appendix-A}.

\section{Finite-dimensional tight-binding dilations for SDE systems}\label{sec: fddil}
The result in \cref{thm:main} relies on exact moment-matching conditions  \eqref{eq:mm-condition} which are usually fulfilled by an infinite-dimensional ancilla space. Many choices are available \cite{li2025linear}, but for practical purposes, we consider the following tight-binding type of dilation, which was derived from an infinite-dimensional dilation using a differential operator $F=p \partial_p + \tfrac12$ on the interval $p\in [0,1].$  

In the continuous setting, this generator $F$ is skew-Hermitian on the  Hilbert space
$\Anc = H_0^1(0,1)$, equipped with the inner product
\[
\langle f,g\rangle = \int_0^1 f^*(p)g(p)\,dp,
\]
and homogeneous boundary condition $f(1)=g(1)=0$. The moment-matching construction, however, is formulated on the larger space $H^1(0,1)$, which contains $H_0^1(0,1)$ as a subspace. For any $\theta>0$, $\theta F$ fulfills the moment condition $\bra{l}(\theta F)^k\ket{r}=1$ for all $k\ge0$ by choosing the right vector as the eigenfunction
$\ket{r}\propto p^\beta$ with $\beta=1/\theta-1/2$, which belongs to $H^1(0,1)$ but not necessarily to $H_0^1(0,1)$ since it does not satisfy the homogeneous boundary condition. The left functional $\bra{l}$ is defined by point evaluation,
\[
\bra{l}f \propto f(p_*),
\]
for some $p_*\in(0,1]$, acting on functions in $H^1(0,1)$, where point evaluation is well defined.

 To obtain a finite-dimensional realization suitable for digital quantum simulation, we partition $[0,1]$ into $M$ intervals with grid points $\{p_i\}_{i=0}^M$. The integration by parts that ensured the skew Hermitian property can be extended to the discrete level using summation by parts (SBP). Following \cite{Strand1994,MattssonNordstrom2004}, 
we let $h_j:=p_{j+1}-p_j$ and define the SBP trapezoid weights
\begin{equation}\label{sbp-wh}
W=\mathrm{diag}(w_0,\dots,w_M),\quad
w_{0}=\tfrac12 h_0,\quad
w_{j}=\tfrac12(h_{j-1}+h_j)\ (1\le j\le M-1),\quad
w_{M}=\tfrac12 h_{M-1}.
\end{equation}
Let $Q$ be the tridiagonal matrix for a centered difference operator with
\begin{equation}\label{sbp-Q}
    (Q)_{j,j+1}=\tfrac12,\quad (Q)_{j+1,j}=-\tfrac12,\quad
Q_{00}=-\tfrac12,\quad Q_{MM}=+\tfrac12,
\end{equation}
and set $D:=W^{-1}Q$. Then the diagonal-norm SBP identity holds:
\[
WD+D^\dagger W
 = 
Q+Q^\dagger
 = 
B:=\mathrm{diag}(-1,0,\ldots,0,1).
\]
This implies that for all grid functions $\bm u, \bm v$,
$\ \langle \bm u,D \bm v\rangle_W+\langle D\bm u,\bm v\rangle_W = u_M v_M - u_0 v_0$ with weighted inner product $\langle \bm x,\bm y\rangle_W:=\bm x^\dag W \bm y$, thus mimicking the integration by parts property, and thus automatically maintains the skew property after the discretization. Specifically, let $P:=\mathrm{diag}(p_0,\ldots,p_M)$ and define the (Hamiltonian) split form
\begin{equation}\label{sbp-Fh}
    F_w  :=  \tfrac12(PD+DP) - \tfrac12 W^{-1}BP,
\qquad
F_h  :=  W^{1/2}F_w  W^{-1/2}.
\end{equation}
The SBP property automatically guarantees that $F_h$ is skew-Hermitian and tridiagonal, while $F_w$ is skew with respect to the $W$–inner product.

Explicitly, the SBP discretization on the geometric grid with $\theta=2$ is given by,
\begin{equation} \label{eq: geometric_grid}
p_j = \exp[- h (M-j)], \quad j=0, 1, \dots, M, \end{equation} where $h > 0$ is a grading parameter (typically $h \approx 1$),
yields a tridiagonal matrix $F_h$ with zeros on the diagonal.  The off-diagonal entries $f_j := (F_h)_{j, j+1} = -(F_h)_{j+1, j}$ take a uniform value in the bulk of the grid, simplifying the implementation:
\begin{equation}\label{eq: fj}
    f_j = \frac{1}{4 \sinh(h /2)} \times 
    \begin{cases} 
    \sqrt{1+e^{-h }}, & j=0, \\
    1, & 1 \le j \le M-2, \\
    \sqrt{1+e^{h }}, & j=M-1.
    \end{cases}
\end{equation}
This nearest-neighbor connectivity allows $F_h$ to be efficiently mapped to a quantum circuit. For example, the operator $iF_h$ corresponds to a hopping Hamiltonian.  It admits a simple 2-local representation using Pauli operators:
\begin{equation}\label{Fh-XY}
    i F_h = -\frac{1}{2} \sum_{j=0}^{M-1} f_j \left( X_j Y_{j+1} - Y_j X_{j+1} \right),
\end{equation}
where $X_j, Y_j$ are the Pauli matrices acting on the $j$-th qubit of the register. This structure is amenable to standard Trotterization or block-encoding techniques on digital quantum processors.

For the eigenvectors $\ket{r} $, we choose $\theta=2$ so $\beta=0$. In addition, we choose the components according to the weights $w_j$: 
\begin{equation}\label{rh}
    \ket{r_h} =  \frac{1}{Z_h} \sum_{j=0}^M \sqrt{w_j} \ket{j}, \quad Z_h = \left( \sum_{j=0}^M w_j \right)^{1/2}.  
\end{equation}
One can show that
\begin{equation}\label{res-err}
   \left( 2 F_h \ket{r_h} - \ket{r_h}  \right) \propto \ket{M}.
\end{equation}
Namely, the residual error is zero for all the interior nodes.  This is due to the SBP discretization. For the evaluation operator, we set it to,
\[
\bra{l_h} =\frac{Z_h}{P_\ast}  \sum_{j=0}^{j_\ast} \sqrt{w_j} \ket{j}, \quad P_\ast =  \sum_{j=0}^{j_\ast}  w_j,
\]
to satisfy $\braket{l_h}{r_h}=1.$  Here $j_\ast$ is an index that we pick to post-select the solution, and its choice will be discussed in the next section.

\begin{remark}
It is necessary to distinguish three levels of the dilation construction used in this paper. First, \cref{thm:main} is formulated abstractly in terms of a moment-matching triple $(F,\ket{r},\bra{l})$ satisfying the condition \eqref{eq:mm-condition}. At this level, no specific realization of the ancillary space or operator $F$ is required.
Second, the differential operator $F = p\partial_p + \frac{1}{2}$ provides an infinite-dimensional realization of such a moment-matching triple. This continuous construction serves as a conceptual model that motivates the finite-dimensional approximation. In this setting, $F$ is understood as an unbounded operator acting on a suitable function space satisfying the stated boundary conditions, while the reference state and evaluation functional are chosen so that the moments in \cref{eq:mm-condition} are well defined.
Finally, the quantum algorithms developed in Sections~\ref{sec:lindblad} and \ref{sec: qtraj} use only the finite-dimensional SBP discretization $F_h$. The matrices $F_h$, together with the corresponding discrete vectors $\ket{r_h}$ and $\bra{l_h}$, constitute the actual objects implemented in the simulation algorithms.
\end{remark}

\subsection{Error Analysis and Stochastic Light-Cone Property}

While the continuous dilation is exact, the finite-dimensional truncation introduces errors due to the discretization of $F$ and the imposition of artificial boundary conditions at $p=1$. To quantify this, we first notice that by introducing a  boundary impurity potential at the edge of the chain:
 \begin{equation}\label{eq:MLC}
    \widehat{F}_h := F_h + \alpha \ketbra{M}{M},\,
    \alpha\!
    := \frac{1}{\theta}
       - \frac{\bra{M}F_h\ket{r_h}}{\bra{M}\ket{r_h}},
 \end{equation}
we have $\theta\widehat{F}_h\ket{r_h}=\ket{r_h}$ is an exact eigenstate. Thus, \( \bra{l_h}\,(\theta\widehat{F}_h)^k\ket{r_h}\!=\!1, \forall k\ge0 \). 
 
The resulting operator, known as a moment-locking closure (MLC) \cite{li2025linear}, in light of \cref{thm:main}, leads to an exact dilation of the SDEs. To leverage this,  we let the corresponding  dilation be, 
\begin{equation}
    \widehat{H}  \coloneqq I_{\mathcal{A}} \otimes H(t) + i \theta \widehat{F}_h \otimes K(t). 
\end{equation}
Similarly, we define the modified drift term, 
\begin{equation}
    \widehat{V}_{0}(t) \coloneqq -i\widehat{H} - \frac{1}{2}\sum_{j=1}^{J} V_j^{\dagger} V_j,
\end{equation}
and let $\ket{\phi_t}$ be the solution of the SDEs 
\begin{equation}\label{eq:dilated-SSE-MLC}
d\ket{\phi_t}
=
\widehat{V}_0(t)\ket{\phi_t}\,dt
+ \sum_{j=1}^J V_j(t)\ket{\phi_t}\,dW_t^j,
\end{equation}
with the same initial condition $\phi(0) = \ket{r_h} \otimes X_0.$ 
Since the dilation using $\widehat{F}_h$ is exact, we have 
\begin{equation}
    \ket{\phi_t} = \ket{r_h} \otimes X_t.
\end{equation}

As a result, the error $\ket{\chi_t} = \ket{\psi_t} - \ket{\phi_t}$ encodes the error 
from the finite-dimensional dilation: using $\braket{l_h}{r_h}=1$, we have 
\[
\bra{l_h} \otimes I \ket{\chi_t} = \bra{l_h} \otimes I \ket{\psi_t} - X_t.
\]
Furthermore, we notice that $\chi_t$
satisfies $\chi(0)=0$ and a driven SSE:
\begin{equation}\label{eq:error-SSE}
    d\ket{\chi_t} = \widetilde{V}_0(t) \ket{\chi_t} dt + \sum_{j=1}^J V_j (t) \ket{\chi_t} dW_t^j + \ket{S_t} dt,
\end{equation}
 where the source term $\ket{S_t} = -\alpha \theta \braket{M}{r_h}  \ket{M} \otimes K(t) X_t$ is localized entirely at the right boundary of the ancilla register (site $M$).

The critical observation is that the error propagates from the boundary into the interior solely through the "hopping" term $F_h \otimes K(t)$ in the drift operator. The noise terms $I \otimes B_j$ are diagonal in the ancilla basis and do not induce spatial transport. This leads to a strong light-cone bound that depends principally on the norm of the dissipative coupling $K(t)$. The following theorem establishes a finite propagation speed that depends explicitly on the tight-binding chain with grading $h $ and the dissipative norm $K_{\max}$. 

\begin{theorem}[Stochastic Light-Cone]\label{thm:light-cone}
Let $\ket{\chi_T}$ be the error state at time $T$ arising from the boundary truncation of the ancilla. Let $j_*$ be an interior ancilla site, and $m = M - j_*$ be the distance from the boundary $j=M$. Let $K_{\max} = \sup_{t} \|K(t)\|$. If the parameters $h$ and $m$ satisfy the condition:
\begin{equation}\label{cfl-bound}
    \varrho := \frac{e \theta K_{\max} T}{4 (M- j_\ast) \sinh(h /2)} < 1,
\end{equation}
then the mean-square error projected onto site $j_*$ decays exponentially with distance:
\begin{equation}
    \mathbb{E}\left[ \| (\bra{j_*} \otimes I) \ket{\chi_T} \|^2 \right] \le \mathcal{C}\frac{\varrho^{2m}}{(1-\varrho)^2}
\end{equation}
where $\mathcal{C}$ is a constant depending on $X(T)$ and grid boundary weights.
\end{theorem}

We defer the proof to \cref{appendix-B}.

Since our bound applies to the mean-square norm of the state error vector $\mathbb{E}[\|\chi_T\|^2]$, this result establishes \emph{strong convergence}  of the dilated quantum simulation. This implies that for any   single noise trajectory, the output state $\ket{\psi_T}$ is physically close to the exact solution, encoding  $X_T$ with high probability, not just consistent in ensemble average.
This result establishes a fundamental speed for error propagation in the stochastic dilation framework. Importantly, this speed depends \emph{only} on the norm of the dissipative operator $K(t)$. It is entirely independent of the magnitude of the Hamiltonian drift $H(t)$ or the strength of the noise $B_j(t)$. As a result, the dimension of the ancilla required scales logarithmically with the precision $\varepsilon $ and linearly with the ``dissipative complexity" $K_{\max} T$. This is analogous to the Lieb-Robinson bounds in many-body physics, where information propagates at a finite velocity determined by the interaction strength.

\begin{remark}
    The moment-locking closure operator $\widehat F_h$ is introduced solely as an auxiliary comparison operator. Since it satisfies $\theta\widehat F_h\ket{r_h}=\ket{r_h},$
    it is no longer skew-Hermitian and therefore does not generate the physical Hamiltonian dilation used in the algorithm. Instead, $\widehat F_h$ defines an exact reference evolution, from which the error equation is derived by comparison with the physical evolution generated by the skew-Hermitian operator $F_h$. Throughout the remainder of the paper, only the Hermitian dilation constructed from $F_h$ is implemented in the quantum algorithms.
\end{remark}

\section{Implementation via Lindblad Simulation Algorithms}\label{sec:lindblad}

A key advantage of the dilation framework is that it enables \emph{ensemble} statistics of the linear
SDE~\eqref{eq:SDE0} to be computed by simulating a \emph{deterministic} quantum master equation, rather
than sampling individual trajectories.  While \cref{thm:main} provides a pathwise embedding
$X_t(\omega)=(\bra{l}\otimes I)\ket{\psi_t(\omega)}$, many quantities of practical interest are
\emph{quadratic} in the state; in particular, for any observable $O$ on $\Hil$,
\begin{equation}
    \mathbb E[X_t^\dagger O X_t]=\tr(\Sigma_t O), \quad \Sigma_t:=\mathbb E\left[X_tX_t^\dagger\right],
\end{equation}
where $\Sigma_t$ is the second-moment (covariance) matrix. Crucially, $\Sigma_t$ evolves deterministically and satisfies a closed second-moment equation.  Our
dilation lifts this deterministic evolution to a quantum master equation on an enlarged Hilbert space, which constitutes a completely-positive and trace preserving (CPTP) dynamic map.
As a result, the estimation of $\tr(\Sigma_t O)$ can be reduced to the estimation of a single
observable on the output of a quantum channel.

Throughout this section we assume $\norm{X_0} = 1$ without loss of generality,
so that $\tr(\Sigma_0) = \norm{X_0}^2 = 1$ and
$\sigma_0 = \ket{X_0}\bra{X_0}$ is already a normalized rank-one state.
Since the second-moment equation and all target observables are linear in
$\Sigma_t$, the result for a general initial condition $X_0$ is recovered by
multiplying the final estimate by $\norm{X_0}^2$.

\subsection{The Dilated Master Equation}

Let $\ket{\psi_t}$ be the solution of the dilated SSE~\eqref{eq:dilated-SSE}.  Define the corresponding density matrix on $\Anc\otimes \Hil$, 
\begin{equation} \label{eq:dilated second-moment}
    \rho_t \;:=\; \mathbb E\!\left[\ket{\psi_t}\!\bra{\psi_t}\right].
\end{equation} 
A direct application of It\^o formula to $\ket{\psi_t}\!\bra{\psi_t}$ yields a Lindblad master equation for
$\rho_t$ (see, e.g., \cite{breuer2002theory}), a universal description of CPTP quantum maps \cite{lindblad1976generators,gorini1976completely}.

\begin{lemma}[Lindblad equation for the dilated second moment] \label{lem:Linblad dilated moment}
The density matrix $\rho_t$ satisfies
\begin{equation}\label{eq:lindblad}
    \frac{d}{dt}\rho_t \;=\; \mathcal{L}_t(\rho_t), \quad  \mathcal{L}_t(\rho_t)\coloneqq
    -i[\widetilde{H}(t), \rho_t]
    + \sum_{j=1}^J \left(
        V_j(t)\rho_t V_j(t)^\dagger
        - \frac{1}{2}\big\{V_j(t)^\dagger V_j(t), \rho_t\big\}
    \right),
\end{equation}
with initial condition $\rho_0=(\ket{r_h}\!\bra{r_h})\otimes \sigma_0$, where
$\sigma_0:=\ket{X_0}\!\bra{X_0}/\|X_0\|^2$ is the normalized rank-one second-moment seed (Thus, $\tr(\rho_0)=1$ due to $\norm{r_h}=1$).  Here
$\widetilde H(t)$ is the Hermitian dilated Hamiltonian \eqref{H-dil} and
$V_j(t)=I_{\mathcal A}\otimes B_j(t)$ are the dilated noise operators.
\end{lemma}

Simulating the Markovian quantum dynamics governed by \eqref{eq:lindblad} is a central primitive in
quantum algorithms.  Early approaches had a polynomial dependence on the precision
\cite{CL17}, while more recent algorithms achieve near-optimal scaling by exploiting higher-order
expansions and block-encoding reductions
\cite{CW17,li2022simulating,ding2024simulating,patel2023wave1}.  For our purposes, it is especially
convenient to use a simulator that outputs a \emph{purification} of $\rho_T$ (e.g.\ \cite{li2022simulating}),
since expectation values of observables can be estimated by standard block-encoding/measurement
routines \cite{rall2020quantum}.

To relate the quadratic statistics of the original SDE \eqref{eq:SDE0} to the density matrix $\rho_T$, recall that 
 $\Sigma_t:=\mathbb E[X_tX_t^\dagger]$.  For the tight-binding dilation, with $\norm{r_h}=1$, one can recover the second moment as follows (up to the light-cone error),
\begin{equation}\label{eq:Sigma-recover-ideal}
\Sigma_t \approx  (\bra{l_h}\otimes I)\,\rho_t\,(\ket{l_h}\otimes I).
\end{equation}
Hence, for any observable $O$ on the system, one has, 
\begin{equation}\label{eq:mu-recover-ideal}
\tr(\Sigma_t O)
\;=\;
\tr\!\Big(\rho_t\,(\Pi_{l_h}\otimes O)\Big),
\qquad \Pi_{l_h}:=\ket{l_h}\!\bra{l_h}.
\end{equation}
Estimating this expectation value via amplitude amplification and block encoding requires $\co(1/\varepsilon )$ rounds of preparations of $\rho_t$ \cite{rall2020quantum}. 

\subsection{Segment-wise evolution and ancilla refresh} \label{refrest-seg}

Due to the finite-dimensional implementation, \eqref{eq:mu-recover-ideal} remains accurate
on each segment up to the controlled light-cone error. One can directly extend \cref{thm:light-cone} to the second moment $\Sigma_t$, as follows,

\begin{proposition}[Light-Cone for the covariance]\label{prop:light-cone-second-moment}
Assume the hypothesis of \cref{thm:light-cone}, i.e., $m=M-j_\ast$, and 
\[
\varrho:=\frac{e\theta K_{\max}T}{4m\sinh(h/2)}<1.
\]
Define the $j_\ast$-localized blocks from  the Lindblad equation \cref{eq:lindblad} as
\[
\rho_T^{(j_\ast)}:=(\bra{j_\ast}\otimes I)\,\rho_T\,(\ket{j_\ast}\otimes I). 
\]
Writing $\gamma:=\braket{j_\ast}{r_h}$, then
the second-moment error on the interior site obeys the trace-norm bound
\begin{equation}\label{eq:lightcone-secondmoment-explicit}
\big\|\rho_T^{(j_\ast)}-|\gamma|^2\Sigma_T\big\|_1
\;\le\;
2|\gamma|\sqrt{\tr(\Sigma_T)}\,
\sqrt{\mathcal{C}\frac{\varrho^{2m}}{(1-\varrho)^2}}
\;+\;
\mathcal{C}\frac{\varrho^{2m}}{(1-\varrho)^2}.
\end{equation}
\end{proposition}
We defer the proof to \cref{sec: lcc}.

The covariance matrix $\Sigma_t$ associated with \cref{eq:SDE0} satisfies a matrix differential equation,
\begin{equation}\label{eq:Sigma}
\frac{\mathrm d}{\mathrm dt}\Sigma_t = \mathcal D_t(\Sigma_t),\quad \Sigma_0 \succeq 0,
\end{equation}
where $\mathcal D_t$ is a generator for a dynamic map that need not be trace-preserving. Without loss of generality, we assume $\tr(\Sigma_0)=1$, since $\Sigma_t$ may be rescaled by the linearity of \cref{eq:SDE0}.

Although the exactness of this dilation is guaranteed for any simulation time $T$, the light-cone analysis indicates that in order to maintain a finite success probability to post-select out $\Sigma_T$ using \cref{eq:Sigma-recover-ideal}, we must choose $T$ such that 
\begin{equation}
 T= \co(K_{\max}^{-1}), \quad K_{\max} \coloneqq \max_{t\in [0,T]} \norm{K(t)}.
\end{equation}
This issue can be circumvented by a segment-wise simulation, and upon the completion of each segment, an oblivious amplitude amplification (OAA) \cite{berry2014exponential,CW17} can be applied to restore the ancilla so that the algorithm can be repeated for the following segment.
Toward this end,  we fix the segment length
\begin{equation}\label{tau-choice}
\tau \coloneqq \co\left(\frac{1}{K_{\max}}\right),\qquad L \coloneqq \left\lceil \frac{T}{\tau}\right\rceil = \co \left( T K_{\max}\right).
\end{equation}
Define the time segments $t_m := m\tau$ for $m=0,1,\dots,L$ (with $t_L \ge T$; one may shorten the last step without affecting the discussion). Let the exact segment evolution from \cref{eq:Sigma} be
\begin{equation}
\mathcal E_m \coloneqq \mathcal T\exp\!\Big(\int_{t_m}^{t_{m+1}}\mathcal D_s\,\mathrm ds\Big),
\end{equation}
so that the ideal final state can be written as the composition
\begin{equation}\label{eq:seg-compose}
\Sigma_T \coloneqq \big(\mathcal E_{L-1}\circ\cdots \circ \mathcal E_1 \circ \mathcal E_0\big)(\Sigma_0).
\end{equation}

Notice that \cref{eq:Sigma} does not necessarily produce a density matrix.  Let us introduce the scalar trace
\[
\lambda_m \coloneqq \tr(\Sigma_{t_m}) \text{ for }m=1,\cdots L,\quad \lambda_0 \coloneqq 1,
\]
and the normalized covariance, which can be regarded as a density matrix, becomes
\[
\sigma_m \;:=\; \frac{\Sigma_{t_m}}{\lambda_m},\qquad \tr(\sigma_m)=1.
\]

To apply a Lindblad simulation algorithm, we define the dilated input at the beginning of segment $m$,
\begin{equation}\label{eq:rho-seg-input}
\rho_{t_m} \;:=\; \ketbra{r_h}\otimes \sigma_m,\qquad \tr(\rho_m)=1,
\end{equation}
which is  the algorithmic working state, without knowing the factors
$\lambda_m$.

Denote $\widetilde{\mathcal E}_m$ as the CPTP segment channel induced by the dilated Lindbladian
\eqref{eq:lindblad} on $[t_m,t_{m+1}]$, and it evolves the system into a \emph{pre-refreshed state},
\begin{equation}\label{eq:rho-pre-refresh}
\rho'_{m+1} \;:=\; \widetilde{\mathcal E}_m(\rho_{t_m}).
\end{equation}

Meanwhile, \cref{thm:main} ensures that the (unnormalized) covariance update over the segment is extracted by the fixed readout
\begin{equation}\label{eq:Sigma-prime-def}
\Sigma'_{m+1}
\;:=\;
(\bra{l_h}\otimes I)\,\rho'_{m+1}\,(\ket{l_h}\otimes I),
\end{equation}
which (up to the controlled light-cone / simulation errors) satisfies
$\Sigma'_{m+1}\approx \mathcal E_m(\sigma_m)$.
We thus define the \emph{segment trace-growth factor} using the unnormalized projector from \cref{eq:mu-recover-ideal}
\begin{equation}\label{eq:gm-def}
g_m \;:=\; \tr(\Sigma'_{m+1})
\;=\;
\tr\!\Big(\rho'_{m+1}\,(\Pi_{l_h}\otimes I)\Big),
\qquad 
\end{equation}
and update the normalized covariance and the scalar trace by
\begin{equation}\label{eq:sigma-lambda-update}
\sigma_{m+1} \;=\; \frac{\Sigma'_{m+1}}{g_m},
\qquad
\lambda_{m+1}\;=\;\lambda_m\,g_m,
\end{equation}
so that the algorithm can proceed to the next time segment. Here
$g_m$ can be estimated by
repeated preparations of $\rho'_{m+1}$ followed by measuring $\Pi_{l_h}$ on the ancilla. 

Iterating \eqref{eq:sigma-lambda-update} yields
\begin{equation}\label{eq:lambda-product}
\lambda_L = \prod_{m=0}^{L-1} g_m, \quad
\Sigma_T = \lambda_L\,\sigma_L, \quad (t_L=T).
\end{equation}
Consequently, for any system observable $O$,
\begin{equation}\label{eq:mu-factorization}
\tr(\Sigma_T O) = \tr(\lambda_L \sigma_L O) = \lambda_L \tr(\sigma_L O) = \left(\prod_{m=0}^{L-1} g_m \right) \tr(\sigma_L O).
\end{equation}

Before proceeding with the same algorithm to the next time segment, another important step is to restore the ancilla to $\ket{r_h}$ so that the initial density matrix for the next segment takes the same form as \cref{eq:rho-seg-input}. By the light-cone property, the restriction of the ancilla to the prefront window $\mathrm{win}=\{0,1,\dots,j_\ast\}$ remains accurate (up to an $\varepsilon $ error) over any segment of length subject to the choice of $\tau$ in \cref{tau-choice}, the overlap \begin{equation} P_\mathrm{win} \coloneqq \sum_{j \in \mathrm{win} } \abs{\braket{j}{r_h}}^2 \end{equation} is a constant lower bound with appropriate choice of $h$ \cite{li2025linear}, and without loss of generality we may assume $P_\mathrm{win} \geq 1/4.$ Let us define the corresponding projector and the truncated ancilla mode, respectively, \begin{equation}\label{eq:PiW} \Pi_{\mathrm{win}} \coloneqq \Big(\sum_{j\in \mathrm{win} }\ket{j}\bra{j}\Big)\otimes I, \quad \ket{r_{\rm win}} \coloneqq \sum_{j\in \mathrm{win} }\ket{j}\bra{j}\ket{r_h}. \end{equation} In particular, $P_\mathrm{win} \geq 1/4$ ensures that the following trace-decreasing CP map succeeds
with constant probability on the relevant states: \begin{equation}\label{eq:refresh-map} \mathcal R_{\mathrm{win}}(\rho) \coloneqq \frac{W_{\mathrm{win}}\, (\Pi_{\mathrm{win}}\rho\,\Pi_{\mathrm{win}}) \,W_{\mathrm{win}}^\dagger} {\tr(\Pi_{\mathrm{win}}\rho)}, \end{equation} where $W_{\mathrm{win}}$ is any fixed isometry on the ancilla register satisfying $W_{\mathrm{win}}\ket{r_{\mathrm{win}}}/\|r_{\mathrm{win}}\|=\ket{r_h}$.

Operationally, one can realize $\mathcal R_{\mathrm{win}}$ either by literal postselection on
$\Pi_{\mathrm{win}}$ (repeat-until-success), or coherently via oblivious amplitude amplification (OAA),
which restores the ancilla without restarting the segment evolution.

More precisely, suppose the Lindblad simulator for segment $m$ outputs a purification of
$\rho'_{m+1}$. There exist an isometry (implemented by the simulator) $U_m$ and an environment
register $E$ such that, for some purification $\ket{\Phi_m}$ of $\rho_m$,
\[
\ket{\psi_{m+1}} := U_m \ket{\Phi_m}\in \Anc\otimes \Hil_{\rm sys}\otimes \mathcal H_E,
\qquad
\tr_E\ket{\psi_{m+1}}\!\bra{\psi_{m+1}}=\rho'_{m+1}.
\]
Define the window projector on the full space (acting trivially on $E$)
\[
\Pi_{\mathrm{win}}^{\rm(full)} := \Pi_{\mathrm{win}}\otimes I_E,
\qquad
q_m := \bra{\psi_{m+1}}\Pi_{\mathrm{win}}^{\rm(full)}\ket{\psi_{m+1}}
      = \tr(\Pi_{\mathrm{win}}\rho'_{m+1}).
\]
By the light-cone property and the choice of $\tau$, we have
$q_m \approx P_{\mathrm{win}}$; in particular, $q_m=\Omega(1)$ (e.g.\ $q_m\gtrsim 1/4$). Here to distinguish the quantum registers, we use $\mathcal H_{\rm sys}$ for the workspace of the SDEs $X_T$, and $\mathcal H_E$ for the additional ancilla for OAA. 

Decompose the post-segment state into its “good” (in-window) and “bad” components:
\[
\ket{\psi_{m+1}} = \ket{G_m}+\ket{B_m},\qquad
\ket{G_m}:=\Pi_{\mathrm{win}}^{\rm(full)}\ket{\psi_{m+1}},\quad
\ket{B_m}:=(I-\Pi_{\mathrm{win}}^{\rm(full)})\ket{\psi_{m+1}},
\]
so that $\braket{G_m}{B_m}=0$ and $\|G_m\|^2=q_m$.
A direct postselection on $\Pi_{\mathrm{win}}$ would succeed with probability $q_m$ and produce the
normalized in-window state $\ket{G_m}/\sqrt{q_m}$.

A more efficient approach is to use OAA, which implements this postselection \emph{coherently} via two reflections:
\[
R_{\mathrm{win}} := I-2\Pi_{\mathrm{win}}^{\rm(full)},
\qquad
R_{r} := 2\bigl(\ket{r_h}\!\bra{r_h}\otimes I_{\rm sys}\otimes I_E\bigr)-I.
\]
Both reflections are ancilla-controlled, and $R_r$ depends only on the fixed reference mode $\ket{r_h}$.
Define the Grover iterate
\[
Q_m := -\,R_{r}\,U_m^\dagger\,R_{\mathrm{win}}\,U_m.
\]
Restricted to the two-dimensional invariant subspace $\mathrm{span}\{\ket{G_m},\ket{B_m}\}$, $Q_m$ acts as a
rotation that amplifies the weight on the “good” subspace. Applying $Q_m$ for
$\Theta(1/\sqrt{q_m})$ iterations boosts the in-window amplitude to $\Theta(1)$; since $q_m=\Omega(1)$,
this requires only $\co(1)$ uses of $U_m$ and $U_m^\dagger$ per segment.

Finally, once the state is supported in $\mathrm{win}$, we deterministically map the truncated ancilla mode
back to the reference by an ancilla-only unitary extension of the isometry. Concretely, let
$\ket{r_{\rm win}}:=\Pi_{\rm win}\ket{r_h}$ and choose any unitary (or isometry extended to a unitary)
$W_{\mathrm{win}}$ on $\Anc$ such that
$W_{\mathrm{win}}\ket{r_{\mathrm{win}}}/\|r_{\mathrm{win}}\|=\ket{r_h}$.
Applying $W_{\mathrm{win}}\otimes I_{\rm sys}\otimes I_E$ completes the refresh and yields an output whose
$\Anc$-marginal is restored to $\ket{r_h}$ (up to the same $\co(\varepsilon )$ light-cone leakage), enabling the
next segment to start again from the canonical form \eqref{eq:rho-seg-input}.

\subsection{Amplitude estimation for the growth factors}

We now discuss how to estimate the growth factors $g_m$.
Write $\beta:=\norm{\ket{l_h}}$ and define the normalized vector and the associated projector,
\[
\ket{\tilde l_h}:=\frac{1}{\beta}\ket{l_h},\qquad \Pi_{\tilde l_h}:=\ket{\tilde l_h}\!\bra{\tilde l_h}.
\]
Here we note that $\beta= \norm{\ket{l_h}} = \sqrt{P_\mathrm{win} } = \Omega(1).$

We define the segment success probability
\begin{equation}\label{eq:pm-def}
q_m \;:=\;\tr\!\Big[\rho'_{t_{m+1}}\,(\Pi_{\tilde l_h}\otimes I)\Big]\in[0,1],
\end{equation}
which can be estimated by measuring the dilation ancilla in the basis
$\{\ket{\tilde l_h},(\ket{\tilde l_h})^\perp\}$ and recording the $\ket{\tilde l_h}$ outcome.
By construction of the recovery functional, the covariance trace update on segment $m$ satisfies
\begin{equation}\label{eq:gm-from-pm}
g_m \;=\;\beta^2\,q_m,
\qquad
\lambda_{m+1}=\lambda_m\,g_m.
\end{equation}

We look for an estimator $\hat g_m$ with relative error
\[
\frac{|\hat g_m-g_m|}{g_m}\le \delta_g,
\qquad \delta_g:=\frac{\varepsilon }{2L},
\]
so that the product $\hat\lambda_L=\prod_{m=0}^{L-1}\hat g_m$ satisfies
$\big|\hat\lambda_L/\lambda_L-1\big|\le e^{L\delta_g}-1 = \co( \varepsilon )$.
Equivalently, we require an additive estimate $\hat q_m$ obeying
\[
|\hat q_m-q_m|\;\le\;\delta_p,\qquad \delta_p:=\delta_g\,q_m=\frac{\varepsilon }{2L\lambda_L }\,q_m.
\]
We further scaled the error by $\lambda_L$ because the expectation $\tr(\Sigma_T O) $ carries a normalizing factor $\lambda_L$. The ability to restore the density matrix without rerunning the previous segment leads to the following complexity bound.

A straightforward method for estimating $g_m$ is to apply AA \cite{rall2020quantum} after each segment. Let $C_{\mathcal L,\tau}$ denote the cost of simulating a single Lindblad segment of duration $\tau$. Consequently, estimating $g_m$ requires $\co(m L/\varepsilon )$ rounds of  $C_{\mathcal L,\tau}$. Because the evolution must be restarted from $t=0$ for each estimation, the total complexity accumulates to $\mathcal{O}(L^3/\varepsilon )$ times $C_{\mathcal L,\tau}$.

It is also possible to avoid restarting the evolution from $t=0$ for each
segment by employing a coherent mean-estimation (amplitude estimation) routine
\cite{Rall2021} applied to the two-outcome measurement
$\{\Pi_{\tilde l_h}\otimes I,\ I-\Pi_{\tilde l_h}\otimes I\}$, followed by
uncomputation to restore the purification of $\rho'_{t_{m+1}}$.

In contrast, the naive approach estimates the growth factor of segment $m$ by
repeatedly preparing $\rho'_{t_{m+1}}$, requiring the evolution to be rerun
from $t=0$ to $t_{m+1}$ for each repetition. The coherent procedure instead
performs the estimation reversibly. After estimating the success probability,
the ancillary workspace is uncomputed, recovering the purification of
$\rho'_{t_{m+1}}$. The restored state can then be used directly as the input to
the next segment, rather than being regenerated from the initial condition.
Consequently, each segment evolution is performed only once (up to the overhead
of amplitude estimation), yielding the complexity stated in
\cref{thm:segwise-lindblad-complexity}.

This latter approach is more efficient, and its overall complexity is summarized as follows,

\begin{theorem}[Segment-wise Lindblad complexity]\label{thm:segwise-lindblad-complexity}
Fix $\tau=\Theta(1/K_{\max})$,  $L=\lceil T/\tau\rceil$ and let $\Lambda_T= \tr(\Sigma_T)$. Assume:
(i) for each segment, there is a Lindblad simulator that implements the CPTP map
$\widetilde{\mathcal E}_m$ for time $\tau$ with cost $C_{\mathcal L,\tau}$ and diamond-norm error at
most $\co(\varepsilon/(\Lambda_T L))$;
(ii) the light-cone/window condition holds so that $P_{\mathrm{win}}\ge 1/4$;
(iii) $\|O\|\le 1$, and we estimate $\nu:=\tr(\sigma_L O)$ from the final normalized state
$\sigma_L$ using a standard expectation-estimation routine for observables \cite{rall2020quantum}, to additive error $\co(\varepsilon/\Lambda_T)$.

Then there is an algorithm that outputs an estimate $\hat\mu$ of
$\mu:=\tr(\Sigma_T O)$ satisfying $|\hat\mu-\mu|\le \varepsilon$ with constant success
probability, using a number of segment-simulation calls scaling as
\[
\widetilde{\mathcal O}\!\left(
\frac{\Lambda_T L}{\varepsilon}
\;+\;
\frac{\Lambda_T L}{\varepsilon}\sum_{m=0}^{L-1}\frac{1}{\sqrt{q_m}}
\right),
\]
up to polylogarithmic factors in $L$, $1/\varepsilon$ and the dimension $N$. Here $q_m$ is defined in \cref{eq:pm-def}. 
\end{theorem}

 The term
$\widetilde \co(\Lambda_T L/\varepsilon)$ comes from estimating the final normalized expectation 
$\nu=\tr(\sigma_L O)$ to additive error $\co(\varepsilon/\Lambda_T)$, which requires repeated segment-wise Lindblad simulations to prepare the purification of $\rho_T$. The other
term accounts for amplitude tracking: to reconstruct the overall scale
$\lambda_L=\prod_{m=0}^{L-1} g_m$ with sufficient accuracy for an $\varepsilon$-additive estimate of
$\mu=\lambda_L\nu$, we estimate each segment success probability
$q_m=\tr[\rho'_{t_{m+1}}(\Pi_{\tilde l_h}\otimes I)]$ to the required precision using coherent
mean-estimation with state restoration.

Near-optimal Lindblad simulation algorithms typically achieve
$C_{\mathcal L,\tau}=\widetilde{\mathcal O}(\|\mathcal L\|\tau)$ (up to polylogarithmic factors),
and in our dilation setting a coarse bound is
\[
\|\mathcal L\|
\leq 
\max_{t\in[0,T]}\Big(\|A(t)\|+\sum_{j}\|B_j(t)\|^2\Big).
\]
In this case, $C_{\mathcal L,\tau} L=  \|\mathcal L\| T. $

\section{Implementation by quantum trajectories}\label{sec: qtraj}

Recall that our dilation scheme reduces the linear SDE~\eqref{eq:SDE0} to an It\^o
stochastic Schr\"odinger equation system
\begin{equation}\label{eq:linSSE}
d\ket{\psi_t}
=
\left(
-i\widetilde H(t)-\frac12\sum_{j=1}^J V_j^\dagger(t)V_j(t)
\right)\ket{\psi_t}\,dt
+\sum_{j=1}^J V_j(t)\ket{\psi_t}\,dW_t^j,
\end{equation}
where $\widetilde H(t)$ is from \cref{H-dil} and is Hermitian, $\{W_t^j\}_{j=1}^J$ are the same independent Wiener processes in the original SDEs \eqref{eq:SDE0}. 

The goal of this section is to address an alternative
simulation task: rather than estimating $\tr(\Sigma_T O)$ via a Lindblad simulator, we aim to
generate a single sample trajectory (pathwise output) at time $T$, i.e.: a quantum state proportional
to the random vector $X_T$.

The operational realization we use is a repeated-interaction scheme, in which the system interacts sequentially to create random paths of the SSEs \cite{lloyd2001engineering,cattaneo2021collision,donvil2023quantum}.

\subsection{Quantum trajectories for linear SSEs: a first-order weak scheme}
\label{sec:weak1}
We start with one noise channel (time index suppressed) and no Hamiltonian term:
\begin{equation}\label{eq:SSE-single}
d\ket{\psi_t}
=
-\frac12 V^\dagger V\,\ket{\psi_t}\,dt
+ V\,\ket{\psi_t}\,dW_t.
\end{equation}
A first-order weak It\^o--Taylor step, also known as the Euler-Maruyama method, is
\[
\psi_{t+\Delta t}
\approx
\Big(I-\tfrac{\Delta t}{2}V^\dagger V + V\,\Delta W_t\Big)\psi_t.
\]
For weak order~1, it suffices to replace $\Delta W_t$ by any random variable whose mean and variance match those of $\Delta W_t$ \cite{kloeden1992numerical}. We use a Rademacher approximation, i.e.:
\begin{equation}\label{eq:xi-rad}
\xi \in \{+\sqrt{\Delta t},-\sqrt{\Delta t}\},
\qquad
\mathbb P(\xi=+\sqrt{\Delta t})=\mathbb P(\xi=-\sqrt{\Delta t})=\frac12,
\end{equation}
so that
\begin{equation}\label{eq:weak-1}
\psi_{t+\Delta t}
\approx
\Big(I-\tfrac{\Delta t}{2}V^\dagger V + \xi\,V\Big)\psi_t.
\end{equation}

A trajectory corresponds to a single run with a 
specific realization of the discrete noise path $\xi$.  Accordingly, we \emph{presample} the Rademacher signs
$s_{n,j}\in\{+1,-1\}$ (equivalently $\xi_{n,j}=s_{n,j}\sqrt{\Delta t}$) \emph{before} running the
quantum circuit, and then \emph{coherently implement} the corresponding conditional update at each
time step.  Operationally, the ancilla measurement serves to select the chosen realization.  

The goal of the remaining subsection is to construct a quantum circuit whose effective action reproduces the weak-order-1 update in \cref{eq:weak-1}. Rather than implementing the non-unitary map directly, we realize it as a postselected branch of a larger unitary evolution acting on the system and a single ancilla qubit. The ancilla measurement encodes the presampled stochastic increment, and the resulting Kraus operator coincides with the Euler–Maruyama weak step up to higher-order terms.

\subsubsection{One-step interaction realization with a single qubit}

We now construct a one-qubit interaction circuit whose effective Kraus operator reproduces the weak-order-1 update \cref{eq:weak-1}. This provides the basic building block for trajectory generation.

Define the anti-Hermitian block generator
\begin{equation}\label{eq:Omega-def}
\Omega
:=
\sqrt{\Delta t}\begin{pmatrix}
0 & -V^\dagger\\
V & 0
\end{pmatrix},
\qquad \Omega^\dagger=-\Omega,
\end{equation}
and the interaction unitary $U:=e^{\Omega}$ acting on a (single-qubit) ancilla and the system.
Applied to $\ket{0}\otimes\ket{\psi}$, a second-order expansion gives
\begin{equation}\label{eq:U-expand}
U\begin{pmatrix}\psi\\ 0\end{pmatrix}
=
\begin{pmatrix}
\big(I-\tfrac{\Delta t}{2}V^\dagger V\big)\psi\\
\sqrt{\Delta t}\,V\psi
\end{pmatrix}
+\co(\dt ^{3/2}).
\end{equation}

  Starting from
$\ket{0}\otimes\ket{\psi}$, apply the interaction unitary $U=e^{\Omega}$. For $s\in\{+1,-1\}$, define the one-qubit unitary $W_s$ by its action
\begin{equation}\label{eq:Ws-def-action}
W_s\ket{0}=\ket{s}_x,\qquad
W_s\ket{1}=\ket{-s}_x,
\end{equation}
where $\ket{\pm}_x=(\ket{0}\pm\ket{1})/\sqrt2$ are the $X$-eigenstates. Postselecting on $\bra{0}$ produces a Kraus branch
\begin{equation}\label{eq:Kraus-preselect-weak1}
M_s \;:=\; \bra{0}\,W_s^\dagger U\,\ket{0},
\qquad
\ket{\psi}\ \mapsto\ M_s\ket{\psi},
\end{equation}
and a second-order expansion of $U$ gives
\begin{equation}\label{eq:psi-preselect-pm}
M_s\ket{\psi}
\ \propto\
\Big(I-\tfrac{\Delta t}{2}V^\dagger V\Big)\ket{\psi}
\ +\ s\sqrt{\Delta t}\,V\ket{\psi}
\ +\ \co(\dt ^{3/2}),
\end{equation}
which matches \eqref{eq:weak-1} exactly with $\xi=s\sqrt{\Delta t}$. Since the target state is $\ket{0},$ we can apply OAA to coherently evolve the system to $\psi_{t_n + \dt}$.

\paragraph{Multiple channels and drift.}
Including the Hamiltonian drift and $J$ channels can be done by operator splitting over one step:
apply $e^{-i \widetilde H(t_n)\Delta t}$ on the system and then apply the above interaction (and $X$-basis
measurement) sequentially for $V_1(t_n),\dots,V_J(t_n)$. This yields the weak order~1 splitting
\begin{equation}\label{split}
    \psi_{t_n+\Delta t}
\approx
\left(\prod_{j=1}^J
\big(I-\tfrac{\Delta t}{2}V_j^\dagger(t_n)V_j(t_n)+\xi_{n,j}V_j(t_n)\big)\right)
e^{-i \widetilde H(t_n)\Delta t}\,\psi_{t_n},
\end{equation}
up to higher-order weak error terms.
\subsubsection{Segmented evolution with ancilla refresh}\label{subsec:qtraj-seg}

As in the Lindblad simulation approach in \cref{refrest-seg}, we exploit the light-cone property:
for times up to $\tau=\Theta(1/K_{\max})$, boundary reflections remain outside the prefront window
$\mathrm{win}=\{0,1,\dots,j_\ast\}$ (up to the controlled light-cone error, achieved by choosing a suitable tight-binding chain length $M$).  We therefore partition
the evolution into segments of length $\tau$ and perform an ancilla refresh at each segment boundary, in analogy with the Lindblad simulations. 
Concretely,  the refresh is implemented as an isometry that (i) flags whether the ancilla
lies in $\mathrm{win}$ (i.e., projects with $\Pi_{\mathrm{win}}$) and (ii) conditionally applies a
fixed ancilla-only isometry $W_{\mathrm{win}}$ mapping
$\ket{r_{\mathrm{win}}}/\|r_{\mathrm{win}}\|\mapsto \ket{r_h}$.  OAA involves two reflections: one about the “good” subspace (equivalently,
$R_{\mathrm{win}}:=I-2\Pi_{\mathrm{win}}$) and one about the prepared flag/ancilla initialization
subspace.  Since the light-cone guarantee implies $P_{\mathrm{win}} \ge\ 1/4$,
for the relevant pre-refresh states, only $\co(1)$ OAA iterations are needed to boost the refresh success probability to a constant.

\subsubsection{Estimating amplitudes }\label{subsec:qtraj-amp}

Because the linear SSE is not norm-preserving pathwise, we represent the unnormalized trajectory state by a normalized quantum state and a classical weight:
\begin{equation}\label{eq:qtraj-weighted-steps}
\ket{\psi_{t_n}} \;=\; \sqrt{\lambda_n}\,\ket{\phi_n},
\qquad \|\ket{\phi_n}\|=1,
\qquad \lambda_n\ge 0.
\end{equation}
Estimating the stepwise factors $g_n$ at every fine step $\Delta t$ is expensive.  Instead, we
\emph{decouple} the estimation timescale from the integration timescale by estimating \emph{products}
of growth factors over coarse blocks.  For notational alignment with the ancilla refresh in the previous section, we take the block
length to be one refresh segment, $\tau$ subject to \eqref{tau-choice}. For simplicity, we choose $\tau= k \dt$. 

Let $\mathcal M_n$ denote the one-step linear map of the chosen weak-$1$ integrator on
$[t_n,t_{n+1}]$ for the presampled increment(s) at step $n$.  Define the \emph{segment map}
\[
\mathcal M_m^{(\tau)}
\;:=\;
\mathcal M_{ mk\dt }\cdots \mathcal M_{ ((m-1) k + 1)  \dt },
\]
and the \emph{segment growth factor}
\begin{equation}\label{eq:qtraj-gm-def}
g_m
\;:=\;
\big\|\mathcal M_m^{(\tau)}\ket{\phi_{(m-1)k}}\big\|^2.
\end{equation}
Thus, $g_m$ is exactly the product of the normalized stepwise growth factors inside segment $m$, but it can be estimated once per segment. In particular,  the $m$th segment admits an implementation
\begin{equation}\label{eq:qtraj-seg-heralded}
U_m\bigl(\ket{0}\otimes \ket{\phi_{(m-1)k}}\bigr)
=
\ket{0}\otimes \bigl(\mathcal M_m^{(\tau)}\ket{\phi_{(m-1)k }}\bigr)
\;+\;
\ket{1}\otimes(\cdots),
\end{equation}
where the flag success probability is exactly $g_m$.  Applying (fixed-point) OAA to \eqref{eq:qtraj-seg-heralded} with the known projector $\ket0\!\bra0$ prepares the normalized post-segment state
\[
\ket{\phi_{mk}^-}
=
\frac{\mathcal M_m^{(\tau)}\ket{\phi_{(m-1)k }}}{\sqrt{g_m}}
\]
with failure probability exponentially small in the number of OAA rounds \cite{brassard2000amplitude,yoder2014fixedpoint,berry2015simulating}. The corresponding overhead is $\widetilde \co(1/\sqrt{g_m})$ uses of $U_m$ and $U_m^\dagger$ (typically constant when $g_m$ fluctuates
around $1$).

\begin{theorem}[Segmented trajectory generation with amplitude tracking]\label{thm:qtraj-seg-cost}
 Let $Y_T$ denote the random
output at time $T$ produced by weak order~1 integrator with \emph{presampled} Rademacher increments, implemented
segment-wise, where each segment contains $k$ inner steps.

Assume:
\begin{enumerate}
\item[(i)] For each segment $m$, it is implemented as,
\[
U_m\bigl(\ket{0}\otimes\ket{\phi_{t_m}}\bigr)
=
\ket{0}\otimes\bigl(\mathcal M_m^{(\tau)}\ket{\phi_{t_m}}\bigr)
+\ket{1}\otimes(\cdots),
\]
whose success probability is
\(
g_m := \|\mathcal M_m^{(\tau)}\ket{\phi_{t_m}}\|^2,
\)
and each call to $U_m$ (or $U_m^\dagger$) costs $C_{{\rm traj},\tau}$. The per-segment simulation error
in the normalized post-segment state is at most $\co(\varepsilon /L)$.
\item[(ii)] The light-cone/window condition holds so that $P_{\mathrm{win}}\ge 1/4$, and the ancilla refresh at each segment boundary
has failure probability and induced state error at most $\co(\varepsilon /L)$ (using $\co(1)$ rounds of fixed-point OAA).
\item[(iii)] Each segment growth factor $g_m$ is estimated in-line (with coherent state restoration) to \emph{relative} accuracy
$\co(\varepsilon /L)$.
\end{enumerate}
Then the algorithm outputs, with constant success probability, a normalized final state
$\ket{\widetilde\phi_T}$ and a scalar $\widetilde\lambda_T$ such that the reconstructed unnormalized trajectory
\[
\widetilde Y_T \;:=\; \sqrt{\widetilde\lambda_T}\,(\bra{l_h}\otimes I)\ket{\widetilde\phi_T}
\]
satisfies the \emph{algorithmic} pathwise error bound
\[
\|\widetilde Y_T - Y_T\| \;=\; \co(\varepsilon ),
\]
and, for any sufficiently smooth test functional $f$ with polynomial growth, the \emph{total weak error} obeys
\[
\big|\E[f(\widetilde Y_T)]-\E[f(X_T)]\big|
\;\le\;
C_T\,\Delta t \;+\; \co(\varepsilon ),
\]
where $C_T$ is the standard weak-$1$ constant.

Moreover, the total number of calls to the segment primitives $\{U_m,U_m^\dagger\}$ scales as
\[
\widetilde \co\!\left(
\sum_{m=0}^{L-1}\frac{1}{\sqrt{g_m}}
\;+\;
\frac{L}{\varepsilon }\sum_{m=0}^{L-1}\frac{1}{g_m}
\right),
\]
and hence the total gate/query complexity is
\[
\widetilde \co\!\left(
C_{{\rm traj},\tau}\left[
\sum_{m=0}^{L-1}\frac{1}{\sqrt{g_m}}
\;+\;
\frac{L}{\varepsilon }\sum_{m=0}^{L-1}\frac{1}{g_m}
\right]
\right),
\]
up to an additional additive overhead $\widetilde \co(L)$ for the $L$ refresh operations (constant-factor in the regime $P_{\rm win}=\Omega(1)$).
\end{theorem}

The first term $\sum_m \widetilde \co(1/\sqrt{g_m})$ is the cost of \emph{state propagation}: to realize the non-unitary segment map
$\mathcal M_m^{(\tau)}$ coherently and output the \emph{normalized} post-segment state
$\ket{\phi_{t_{m+1}}^-}\propto \mathcal M_m^{(\tau)}\ket{\phi_{t_m}}$ without restarting, we apply fixed-point OAA to the segment heralding flag, which costs $\widetilde \co(1/\sqrt{g_m})$ uses of $\{U_m,U_m^\dagger\}$ on segment $m$.
The second term $\frac{L}{\varepsilon }\sum_m \widetilde \co(1/g_m)$ is the cost of \emph{weight tracking}.

\subsection{Second-Order Weak Scheme via Weak Measurement}\label{sec:Magnus-3pt-weak}

We develop a \emph{weak order-$2$} one-step approximation for quantum trajectories of the linear
It\^o SSE with \emph{scalar noise}
\begin{equation}\label{eq:linSSE-AV}
d\ket{\psi_t}=A(t)\ket{\psi_t}\,dt+V(t)\ket{\psi_t}\,dW_t,
\qquad t\in[0,T].
\end{equation}
Again, we consider a single noise channel, where 
$A(t)=-\tfrac12 V^\dagger(t)V(t)$.  The coherent term from $\widetilde H$ and multiple jump operators can be treated by generalizing \cref{split} to a symmetric trotter splitting.

We emphasize that the goal is
\emph{weak} accuracy: the one-step map should reproduce expectations of smooth functionals up to
$\co(\dt^3)$ \emph{local} weak error (and hence $\co(\Delta t^2)$ \emph{global} weak error).

Besides the Brownian increment $\Delta W_n:=W_{t_{n+1}}-W_{t_n}$, weak order~$2$ requires the second
It\^o integral
\begin{equation}\label{eq:DeltaZ-def-qtraj}
\Delta Z_n:=\int_{t_n}^{t_{n+1}}(W_s-W_{t_n})\,ds
=\int_{t_n}^{t_{n+1}}\!\!\int_{t_n}^{s} dW_r\,ds,
\end{equation}
and the related iterated integrals \cite{kloeden1992numerical}
\begin{equation}\label{eq:I11-I10-def-qtraj}
I_{11,n}:=\int_{t_n}^{t_{n+1}}\!\!\int_{t_n}^{s} dW_r\,dW_s
=\frac12\big((\Delta W_n)^2-\Delta t\big),
\;
I_{10,n}:=\int_{t_n}^{t_{n+1}}\!\!\int_{t_n}^{s} dr\,dW_s
=\Delta t\,\Delta W_n-\Delta Z_n.
\end{equation}
The identity for $I_{11,n}$ follows from It\^o isometry, while the relation for $I_{10,n}$ is a
direct It\^o integration by parts.

A weak It\^o--Taylor expansion of order $2.0$ \cite{kloeden} gives the one-step local expansion
\begin{equation}\label{eq:weak2-ito-taylor-tn-clean}
\begin{aligned}
\ket{\psi_{n+1}}
&=
\Big[
I
+ \Delta t\,A(t_n)
+ V(t_n)\,\Delta W_n
+ V(t_n)^2\,I_{11,n} \\
&
+ \big(\dot V(t_n)+V(t_n)A(t_n)\big)\,I_{10,n}
+ \big(A(t_n)V(t_n)\big)\,\Delta Z_n \\
& + \frac{\Delta t^2}{2}\,\big(\dot A(t_n)+A(t_n)^2\big) 
\Big]\ket{\psi_n}
+\co_{\rm w}(\Delta t^3),
\end{aligned}
\end{equation}
where $\co_{\rm w}(\Delta t^3)$ denotes a remainder whose contribution to \emph{weak} local error is
$\co(\Delta t^3)$ under standard boundedness and regularity assumptions on $A(\cdot),V(\cdot)$. In
particular, for smooth test functionals $f$ one obtains a global weak error bound of the form
\begin{equation}\label{eq:weak2-bound-clean}
\big|
\mathbb E[f(\ket{\psi_T})]-\mathbb E[f(\ket{\widehat\psi_N})]
\big|
\le
C_T\,\Delta t^{2},
\end{equation}
with $C_T$ depending on $T$ and on uniform bounds for $A,V$ and the derivatives required by the weak-$2$
theory \cite{kloeden}.

To streamline subsequent circuit constructions, we evaluate coefficients at the midpoint
$t_{n+\frac12}:=t_n+\Delta t/2$:
\begin{equation}\label{eq:mid-defs-qtraj}
A_{\rm mid}:=A(t_{n+\frac12}),\qquad V_{\rm mid}:=V(t_{n+\frac12}),\qquad
\dot V_{\rm mid}:=\dot V(t_{n+\frac12}).
\end{equation}
For example, for smooth $A(\cdot)$ we have the Taylor relation
\[
A_{\rm mid}=A(t_n)+\frac{\Delta t}{2}\dot A(t_n)+\co(\Delta t^2),
\]
so the combination $\Delta t\,A(t_n)+\frac{\Delta t^2}{2}\dot A(t_n)$ appearing in
\eqref{eq:weak2-ito-taylor-tn-clean} is \emph{absorbed} into $\Delta t\,A_{\rm mid}$ up to
$\co(\Delta t^3)$. Moreover, replacing $A(t_n)$ by $A_{\rm mid}$ in the quadratic term $A(t_n)^2$
changes it only by $\co(\Delta t)$, hence contributes $\co(\Delta t^3)$ after multiplication by
$\Delta t^2$. Therefore, \eqref{eq:weak2-ito-taylor-tn-clean} can be rewritten (without an explicit
$\dot A$ term) as the midpoint weak-$2$ expansion
\begin{equation}\label{eq:weak2-ito-taylor-mid-clean}
\begin{aligned}
\ket{\psi_{n+1}}
=
\Big[
&I
+ \Delta t\,A_{\rm mid}
+ V_{\rm mid}\,\Delta W_n
+ V_{\rm mid}^2\,I_{11,n} \\
&\quad
+ \big(\dot V_{\rm mid}+V_{\rm mid}A_{\rm mid}\big)\,I_{10,n}
+ \big(A_{\rm mid}V_{\rm mid}\big)\,\Delta Z_n
+ \frac{\Delta t^2}{2}\,A_{\rm mid}^2
\Big]\ket{\psi_n}
\;+\;\co_{\rm w}(\Delta t^3).
\end{aligned}
\end{equation}
This is the form we will discretize and then implement via weak measurement.

The pair $(\Delta W_n, \Delta Z_n)$ forms a centered Gaussian vector with the following covariance structure:$$\text{Cov}\begin{pmatrix} \Delta W_n \\ \Delta Z_n \end{pmatrix}
= \mathbb{E}\left[ \begin{pmatrix} \Delta W_n \\ \Delta Z_n \end{pmatrix} \begin{pmatrix} \Delta W_n & \Delta Z_n \end{pmatrix} \right]
= \begin{pmatrix} \Delta t & \frac{1}{2}\Delta t^2 \\ \frac{1}{2}\Delta t^2 & \frac{1}{3}\Delta t^3 \end{pmatrix}.$$
 A convenient representation is obtained by Cholesky factorization and
introducing independent standard normal variables $\xi_{1,n},\xi_{2,n}\sim\mathcal N(0,1)$ and setting
\begin{equation}\label{eq:DW-DZ-xi12-gauss}
\Delta W_n=\sqrt{\Delta t}\,\xi_{1,n},\;
\Delta Z_n=\frac{\Delta t^{3/2}}{2}\Big(\xi_{1,n}+\frac{1}{\sqrt3}\xi_{2,n}\Big).
\end{equation}
As a result, 
\begin{equation}\label{eq:I11-xi1}
\begin{aligned}
    I_{10,n}& =\Delta t\,\Delta W_n-\Delta Z_n=\frac{\Delta t^{3/2}}{2}\Big(\xi_{1,n}-\frac{1}{\sqrt3}\xi_{2,n}\Big), \\ 
    I_{11,n}&=\frac12\big((\Delta W_n)^2-\Delta t\big)
=\frac{\Delta t}{2}\big(\xi_{1,n}^2-1\big).
\end{aligned}
\end{equation}
Substituting \eqref{eq:DW-DZ-xi12-gauss}--\eqref{eq:I11-xi1} into \eqref{eq:weak2-ito-taylor-mid-clean}
and collecting powers of $\Delta t$ yields the compact local form
\begin{equation}\label{eq:weak2-local-xi12}
\begin{aligned}
\ket{\psi_{n+1}} & =F^{(2)}_n  \ket{\psi_{n} },\\
F^{(2)}_n
& \coloneqq 
I
+ \Delta t\,A_{\rm mid} +  \frac{\Delta t^2}{2}\,A_{\rm mid}^2 
+ \sqrt{\Delta t}\,\xi_{1,n}\,V_{\rm mid}
+ \frac{\Delta t}{2}\big(\xi_{1,n}^2-1\big)V_{\rm mid}^2 \\
&\quad
+ \frac{\Delta t^{3/2}}{2}\,\xi_{1,n}\,\big(A_{\rm mid}V_{\rm mid}+\dot V_{\rm mid}+V_{\rm mid}A_{\rm mid}\big)
+ \frac{\Delta t^{3/2}}{2\sqrt3}\,\xi_{2,n}\,\big(A_{\rm mid}V_{\rm mid}-\dot V_{\rm mid}-V_{\rm mid}A_{\rm mid}\big).
\end{aligned}
\end{equation}
It is convenient to package the $\Delta t^{3/2}$ operators as
\begin{equation}\label{eq:BC-mid-clean}
\begin{aligned}
B_{\rm mid}:=&\frac12\big(A_{\rm mid}V_{\rm mid}+\dot V_{\rm mid}+V_{\rm mid}A_{\rm mid}\big),\\
C_{\rm mid}:=&\frac{1}{2\sqrt3}\big(A_{\rm mid}V_{\rm mid}-\dot V_{\rm mid}-V_{\rm mid}A_{\rm mid}\big).
\end{aligned}
\end{equation}

On the other hand, for weak order~$2$ it is sufficient to replace the Gaussian $\xi_{1,n},\xi_{2,n}$ by discrete random
variables that match moments up to order~$4$. We therefore use the Kloeden--Platen three-point law:
\begin{equation}\label{eq:KP-3pt-weak2}
\xi_{k,n}\in\{0,\pm\sqrt3\},\qquad
\mathbb P(\xi_{k,n}=0)=\frac23,\quad
\mathbb P(\xi_{k,n}=\pm\sqrt3)=\frac16,
\qquad k\in\{1,2\},
\end{equation}
so that $\mathbb E[\xi]=0$, $\mathbb E[\xi^2]=1$, $\mathbb E[\xi^3]=0$, $\mathbb E[\xi^4]=3$.
Under \eqref{eq:KP-3pt-weak2}, the squared variable $\xi_{1,n}^2$ takes only two values:
\[
\xi_{1,n}^2\in\{0,3\},\qquad \mathbb P(\xi_{1,n}^2=0)=\frac23,\quad \mathbb P(\xi_{1,n}^2=3)=\frac13.
\]
For circuit design it is convenient to treat this as an explicit ``control variable'' for the 
$V_{\rm mid}^2$ term. We therefore introduce a mean-zero random variable,
\begin{equation}\label{eq:xi3-def}
\xi_{3,n}:=\xi_{1,n}^2 -1 \in\{-1,2\}.
\end{equation}
Note that $\xi_{3,n}$ is \emph{centered} and \emph{pairwise uncorrelated} with $\xi_{1,n}$:
\(
\mathbb E[\xi_{3,n}]=0
\)
and
\(
\mathbb E[\xi_{1,n}\xi_{3,n}]=\mathbb E[\xi_{1,n}^3]-\mathbb E[\xi_{1,n}]=0
\),
although $(\xi_{1,n},\xi_{3,n})$ are not independent.

With this notation, the weak-$2$ one-step update can be written in a compact form:
\begin{equation}\label{eq:weak2-local-xi123}
F^{(2)}_n
=I + \Delta t\,A_{\rm mid} + \frac{\Delta t^2}{2}\,A_{\rm mid}^2 + \sqrt{\Delta t}\,\xi_{1,n}\Big( V_{\rm mid} 
+ \Delta t\,B_{\rm mid} \Big)+ \Delta t^{3/2}\,\xi_{2,n}\,C_{\rm mid}
+ \frac{\Delta t}{2}\,\xi_{3,n}\,V_{\rm mid}^2,
\end{equation}
where $B_{\rm mid},C_{\rm mid}$ are given in \eqref{eq:BC-mid-clean}. \cref{eq:weak2-local-xi123} is the target algebraic form for our second-order weak-measurement step.

\medskip

The following theorem, as motivated by the construction in \cite{ding2024simulating}, summarizes the implementation of  \cref{eq:weak2-local-xi123} using a repeated interaction scheme with two ancilla qubits. 

\begin{theorem}[Two-qubit weak-measurement realization of the weak-$2$ step]\label{thm:weak2-2q-ancilla}
Consider a two-qubit ancilla with basis \(\{\ket{00},\ket{10},\ket{01},\ket{11}\}\) and define the
anti-Hermitian generator
\begin{equation}\label{eq:Kweak2-one-dilation}
\Omega_n
:=
\sum_{\alpha\in\{10,01,11\}}
\Big(\ket{\alpha}\!\bra{00}\otimes G_{\alpha,n}-\ket{00}\!\bra{\alpha}\otimes G_{\alpha,n}^\dagger\Big),
\qquad U_n:=e^{\Omega_n}.
\end{equation}
Let \(V_{\rm mid},A_{\rm mid},B_{\rm mid},C_{\rm mid}\) be as in
\eqref{eq:mid-defs-qtraj} and \eqref{eq:BC-mid-clean}, with the noise-block choice
\(A_{\rm mid}=-\tfrac12 V_{\rm mid}^\dagger V_{\rm mid}\).
Choose
\begin{equation}\label{eq:G123-choice-rev}
\begin{aligned}
G_{10,n}\equiv G_{1,n}
&:=\sqrt{\Delta t}\,V_{\rm mid}
+\Delta t^{3/2}\Big(B_{\rm mid}+\tfrac16 V_{\rm mid}V_{\rm mid}^\dagger V_{\rm mid}\Big),\\
G_{01,n}\equiv G_{2,n}
&:=\frac{\Delta t}{\sqrt2}\,V_{\rm mid}^2,\\
G_{11,n}\equiv G_{3,n}
&:=\Delta t^{3/2}\,C_{\rm mid}.
\end{aligned}
\end{equation}

For each presampled pair \((\xi_{1,n},\xi_{2,n})\) drawn from \eqref{eq:KP-3pt-weak2}, set
\(\xi_{3,n}:=\xi_{1,n}^2-1\) and define the ancilla state
\begin{equation}\label{eq:m-xi12-def}
\ket{m(\xi_{1,n},\xi_{2,n})}
\;:=\;
\frac{1}{\alpha(\xi_{1,n},\xi_{2,n})}\Big(
\ket{00}
+\xi_{1,n}\ket{10}
+\frac{\xi_{3,n}}{\sqrt2}\ket{01}
+\xi_{2,n}\ket{11}
\Big),
\qquad \alpha(\xi_{1,n},\xi_{2,n})>0.
\end{equation}
Let the post-selected effective one-step map be
\begin{equation}\label{eq:Keff-one-dilation}
\widetilde F_n(\xi_{1,n},\xi_{2,n})
\;:=\;
\frac{\bra{m(\xi_{1,n},\xi_{2,n})}\,U_n\,\ket{00}}{\braket{m(\xi_{1,n},\xi_{2,n})}{00}}.
\end{equation}
Then, for any system state \(\ket{\psi_n}\),
\begin{equation}\label{eq:weak2-expansion-one-dilation}
\widetilde F_n(\xi_{1,n},\xi_{2,n})\ket{\psi_n}
=
F^{(2)}_n\ket{\psi_n}
+\co_{\rm w}(\Delta t^3),
\end{equation}
i.e., the update reproduces the weak It\^o--Taylor step
\eqref{eq:weak2-local-xi123} with local weak error \(\co(\Delta t^3)\).
\end{theorem}
We present the proof in \cref{appendix-order2}.

\paragraph{Multiple channels and drift.}
Including the Hamiltonian drift and $J$ channels can be done by operator symmetric splitting over one step:
\begin{equation}\label{second-order split}
    \psi_{t_n+\Delta t}\approx e^{-i \widetilde H(t_n)\Delta t/2}\left(\prod_{j=1}^J F_{n,j}^{(2)}\left(\frac{\dt}{2}\right)\right)\left(\prod_{j=0}^{J-1}F_{n,J-j}^{(2)}\left(\frac{\dt}{2}\right)\right)
e^{-i \widetilde H(t_n)\Delta t/2}\,\psi_{t_n},
\end{equation}
up to higher-order weak error terms. Here $F_{n,j}^{(2)}$ is the corresponding $F_n^{(2)}$ for each noise channel $V_j$ in \cref{eq:weak2-local-xi123}.

\section{Numerical Simulations}\label{sec: numsim}

In this section, we present numerical results that verify our quantum simulation framework through three representative tests.

In the first numerical test, we demonstrate the recovery of the non-unitary evolution operator for linear SDEs via moment-matching dilation in \cref{thm:main}. We consider the example in \cite[Example 5.2]{buckwar2010towards}, a three-dimensional SDE system
\begin{equation}\label{eq:example}
    d\mat{X_1(t) \\ X_2(t) \\ X_3(t)} =
    \mat{-1 & 10 & 0 \\ 0 & -1 & 10 \\ 0 & 0 & -1}
    \mat{X_1(t) \\ X_2(t) \\ X_3(t)}\,dt
    + \sum_{j=1}^3
    \mat{\frac{\sigma}{\sqrt{3}} & 0 & 0 \\ 0 & \frac{\sigma}{\sqrt{3}} & 0 \\ 0 & 0 & \frac{\sigma }{\sqrt{3}} }
    \mat{X_1(t) \\ X_2(t) \\ X_3(t)}\,dW_t^j.
\end{equation}

\begin{figure}
\begin{subfigure}{0.5\textwidth}
\includegraphics[width=0.9\linewidth, height=6cm]{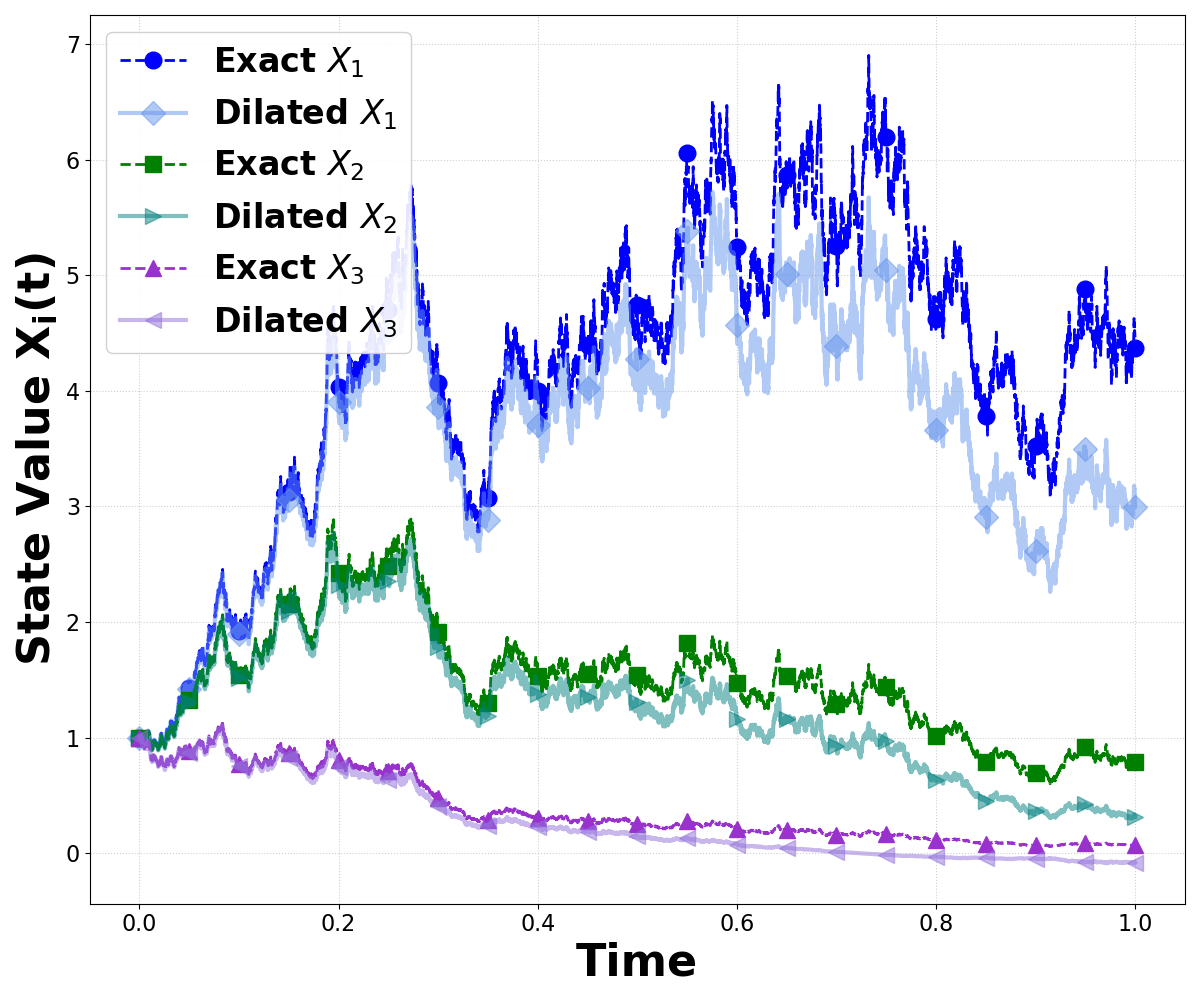}
\caption{$p_* = 0.4$.}
\label{fig:p04}
\end{subfigure}
\begin{subfigure}{0.5\textwidth}
\includegraphics[width=0.9\linewidth, height=6cm]{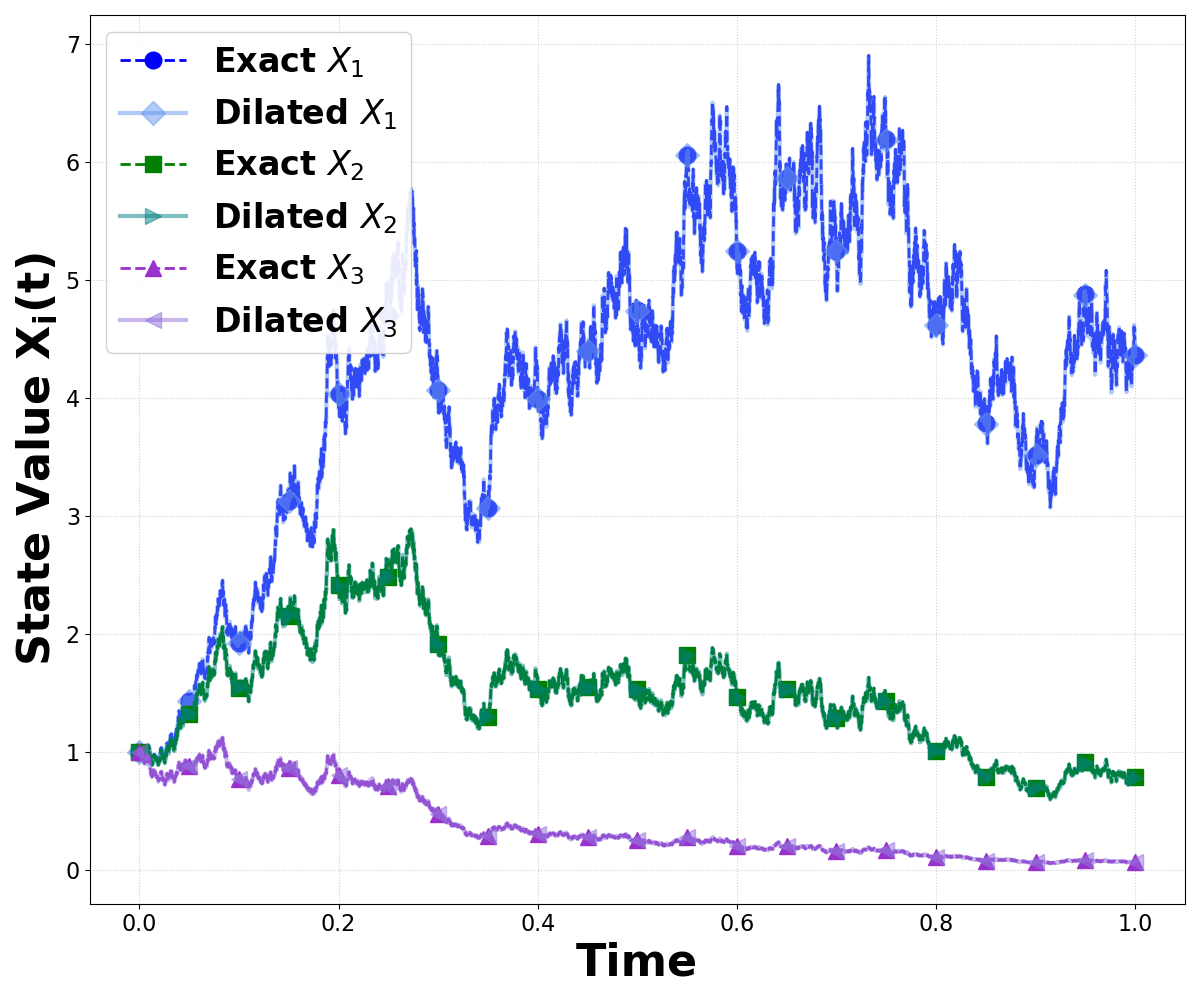}
\caption{$p_* = 0.1$.}
\label{fig:p01}
\end{subfigure}
\caption{Moment-matching dilation for \cref{eq:example} with $\sigma=1$. Both panels compare trajectories from \cref{eq:example} with the projected dilated SSE \cref{eq:dilated-SSE}, driven by the same Brownian motion.
Choosing an evaluation site closer to the origin improves the accuracy and extends the time horizon over which the projected dynamics remain reliable.
In particular, the approximation deteriorates earlier for $p_*=0.4$ (\cref{fig:p04}) than for $p_*=0.1$ (\cref{fig:p01}). }
\label{fig:diff_p}
\end{figure}

We apply the dilation using \cref{eq: fj,Fh-XY}, and we examine the choice of $j_\ast$ used to post-select the solution, as highlighted by the condition \eqref{cfl-bound} in \cref{thm:light-cone}. Specifically, we simulate the dilated SDE \eqref{eq:dilated-SSE} corresponding to \cref{eq:example} and then apply the localized readout
\[
\frac{1}{\braket{j_\ast}{r_h}}\;(\bra{j_\ast}\otimes I)
\]
to $\ket{\psi_t}$ to extract an approximation of $X_t$.
\cref{fig:diff_p} shows that this localized projection yields a more accurate approximation over a longer time interval when $p_*=p_{j_\ast}$ is closer to the origin, consistent with the light-cone property encoded in \eqref{cfl-bound}.

\smallskip
In the second numerical test, we verify the expected error scaling of our second-order weak trajectory scheme \eqref{eq:weak2-local-xi12} by two examples. We pick $B\in\mathbb{R}^{3\times 3}$ from a randomly generated matrix,
\[
B = \mat{-0.79312248 & 0.24057128 & -1.89632635 \\
 1.39577171 & 0.63829474 & -0.29204749 \\
 -0.31194933 & 0.30383537 & -0.2676603}.
\]
We then consider the linear SDE,
\begin{equation} \label{eq: example_weak}
    dX_t = -\frac{1}{2}B^\dagger B\, X_t \,dt + B X_t\,dW_t, \qquad X(0) = \mat{1 & 1 & 1}^T,
\end{equation}
which was discussed in \cref{sec:weak1,sec:Magnus-3pt-weak}.

We do another numerical verification with two noise channels and a small drift term,
\begin{equation} \label{eq: example_weakgen}
    dX_t = A X_t dt + B_1 X_t dW_t^1 + B_2 X_t dW_t^2,
\end{equation}
where 
\[ B_1 = \mat{-1.9250645 & -3.01879523 & 1.15446315 \\
  1.56278191 & 1.32951737 & -0.4295925 \\
 -0.07288832 & -0.29154696 & 1.36824268 }, \]
 \[
 B_2 = \mat{0.87196735 & 0.05837772 & 0.70812363 \\
 1.36030603 & 0.37452505 & -1.08551679 \\
 -0.07479491 & -0.42665655 & -1.59184729
 },\]
 \[A = 0.05 \mat{1 & 0 & 0 \\ 0 & -0.5 & 0 \\ 0 & 0 & 0.25}-\frac{1}{2}B_1^\dag B_1 - \frac{1}{2}B_2^\dag B_2. \]

\begin{figure}
\begin{subfigure}{0.5\textwidth}
\includegraphics[width=0.9\linewidth, height=6cm]{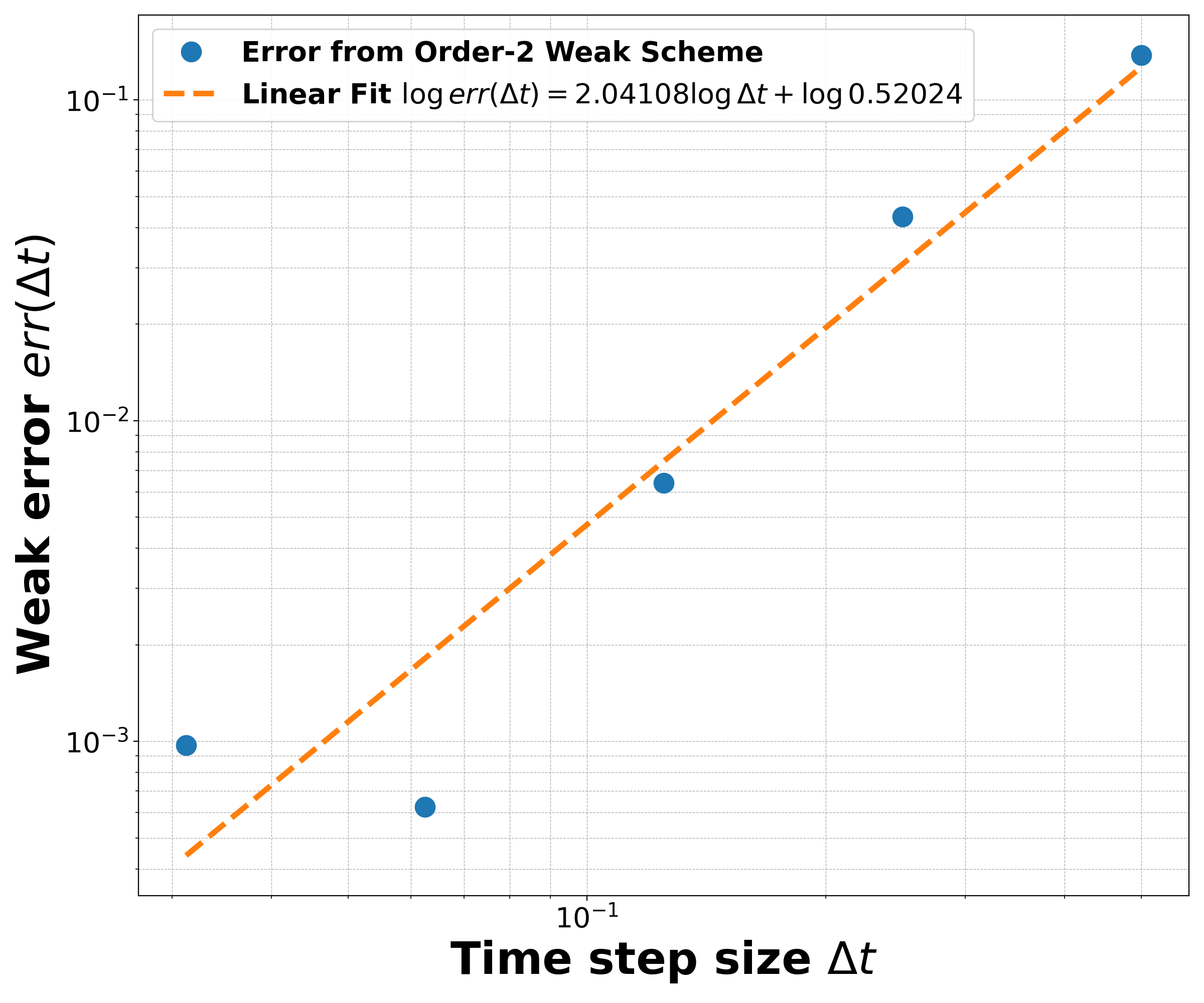}
    \caption{The weak convergence performance for single channel \cref{eq: example_weak}.}
    \label{fig:2nd_order_err}
\end{subfigure}
\begin{subfigure}{0.5\textwidth}
\includegraphics[width=0.9\linewidth, height=6cm]{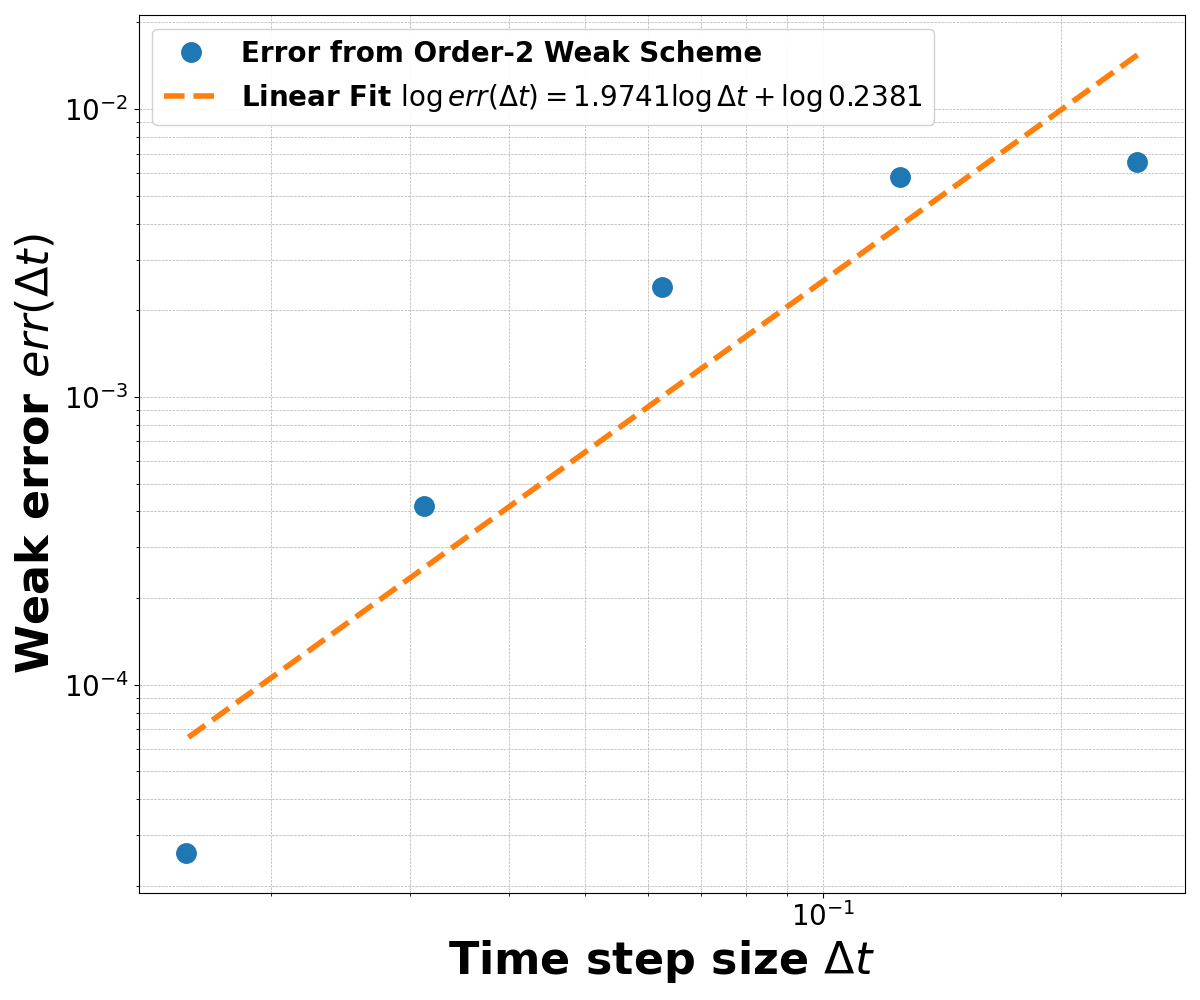}
\caption{The weak convergence performance for two noise channels with a small drift term \cref{eq: example_weakgen}.}
\label{fig:2nd_Order_errgen}
\end{subfigure}
\caption{Weak error $\mathrm{err}(\Delta t) = \bigl|\mathbb{E}[f(Y_{T,\dt}) - \mathbb{E}[f(X_T)]]\bigr|$ for the second-order scheme applied to \cref{eq: example_weak} with $f(\bm x)=\cos(x_1+x_2+x_3^2)$ at $T=1.0$. The number of samples is $N_{\operatorname{samp}} = 5000000$. In \cref{fig:2nd_Order_errgen}, we implement the symmetric splitting for the Hamiltonian term $e^{-i\widetilde{H}\Delta t}$, \cref{second-order split}. In the dilation on the geometric grid \cref{eq: geometric_grid}, we use M=50, and $h=2$.}
\label{fig:2nd_order_weak}
\end{figure}

To assess weak convergence, we test the smooth function $f(\bm x)=\cos\!\left(x_1+x_2+x_3^2\right)$ and define the weak error at final time $T$ by
\[
\mathrm{err}(\Delta t)\coloneqq \bigl|\mathbb{E}[f(Y_{T,\dt}) - \mathbb{E}[f(X_T)]]\bigr|,
\]
where $Y_{T,\dt}$ is computed by the proposed second-order weak scheme \eqref{eq:weak2-local-xi123} with step size $\Delta t$, and $X_{T}$ is a reference solution computed with the Euler--Maruyama method using a much smaller step $\delta t = 2^{-14}$ over $10^7$ realizations. \cref{fig:2nd_order_weak} plots $\mathrm{err}(\Delta t)$ versus $\Delta t$ on a log--log scale. A linear fit of the data reveals an empirical rate close to $2$, confirming the global weak second-order convergence predicted by the analysis.

\smallskip
In the last numerical test, we validate the recovery of the second moment for a stochastic PDE (SPDE) after dilation and transformation into a Lindblad equation (\cref{lem:Linblad dilated moment}). Specifically, we consider an It\^{o} SPDE, a stochastic advection-diffusion-reaction equation in \cite[Example 4.2]{zhang2015wiener}, on $(0,T]\times(0,2\pi)$ with periodic boundary conditions:
\begin{equation} \label{eq:example_Magnus_Linblad}
\begin{aligned}
    du &= \left[ \left( \varepsilon  + \frac{1}{2}\sigma_1^2 \cos^2(x) \right) \partial_x^2 u
    + \left( \beta \sin(x) - \frac{1}{4}\sigma_1^2 \sin(2x) \right) \partial_x u \right] dt \\
    & \quad + \sigma_1 \cos(x)\partial_x u \, dW_t^1 + \sigma_2 u \, dW_t^2,
    \qquad u(x,0) = \sin(x).
\end{aligned}
\end{equation}
Let $D_1, D_2\in\mathbb{R}^{N\times N}$ be the first- and second-order finite difference discretizations using central differences on $N$ grid points . Denoting the semi-discrete solution vector by $X_t \in\mathbb{R}^N$,
the SPDE reduces to a linear system of multiplicative-noise SDEs,
\begin{equation} \label{eq:example_ito}
    dX_t = AX_t \,dt + B_1X_t \,dW_t^1 + B_2X_t \,dW_t^2,
\end{equation}
where
\begin{align}
    A =& \operatorname{diag}\!\left(\varepsilon +\tfrac{1}{2}\sigma_1^2\cos^2(x)\right)D_2 
     + \operatorname{diag}\!\left(\beta \sin(x)-\tfrac{1}{4}\sigma_1^2\sin(2x)\right)D_1, \\
B_1=& \operatorname{diag}(\sigma_1 \cos(x))D_1, \\
B_2 =& \sigma_2 I.
\end{align}

\begin{figure}
\begin{subfigure}{0.5\textwidth}
\includegraphics[width=0.9\linewidth, height=6cm]{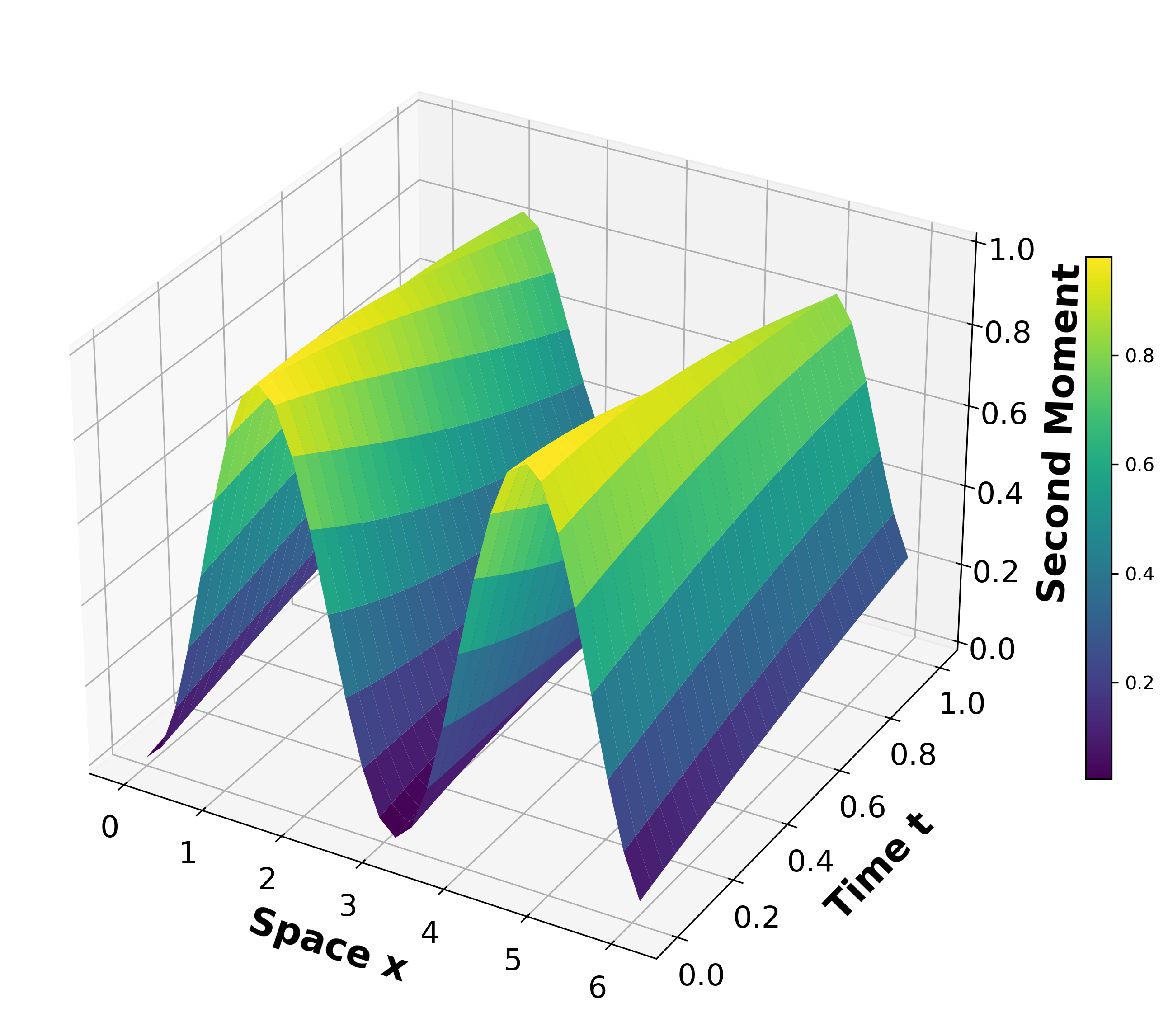}
\caption{Reference second moment obtained from the closed second-moment equation in \cref{eq:spde-second-moment}.}
\label{fig:SPDE_exact_var}
\end{subfigure}
\begin{subfigure}{0.5\textwidth}
\includegraphics[width=0.9\linewidth, height=6cm]{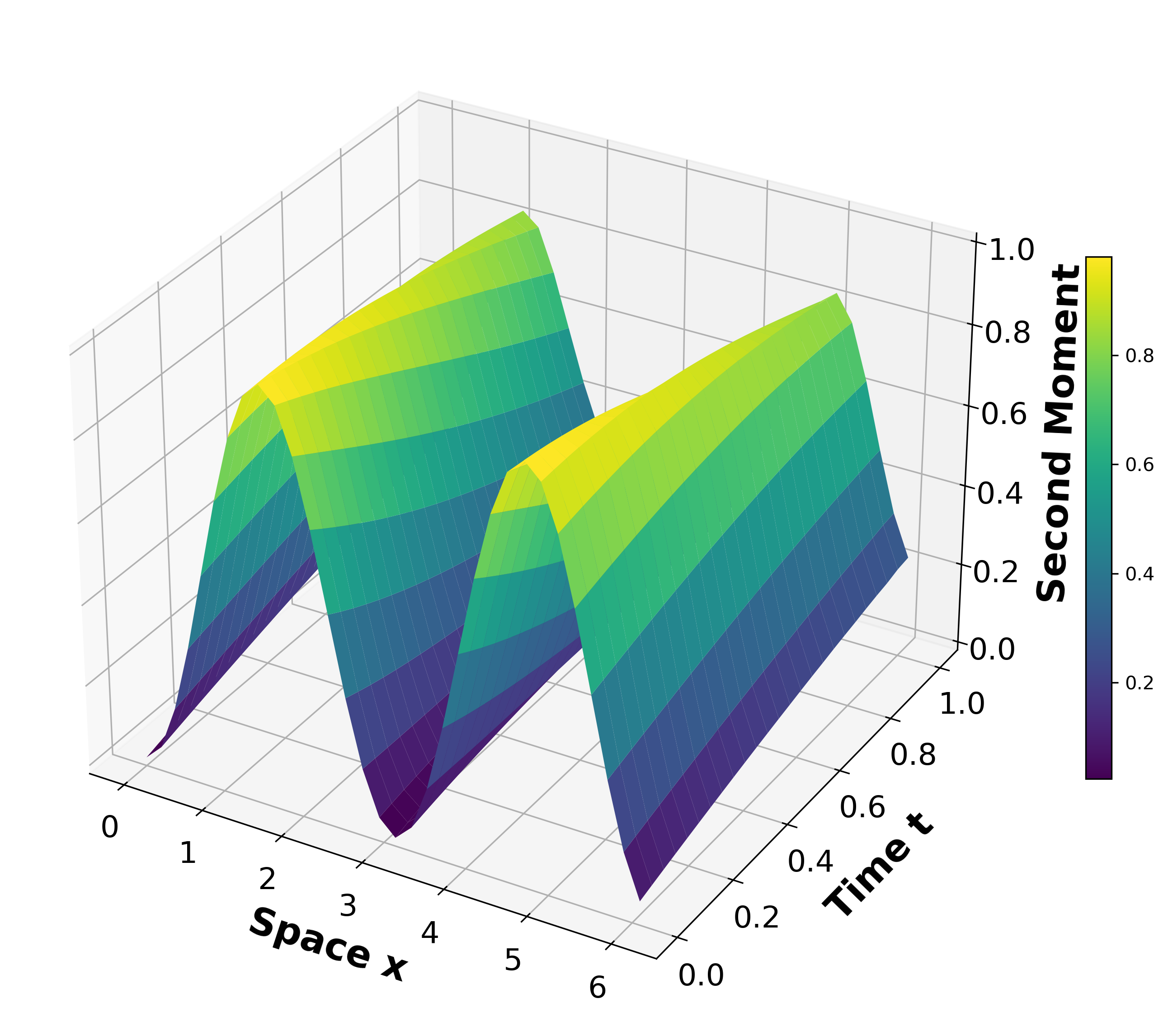}
\caption{Recovered second moment using the dilation-based Lindblad simulation.}
\label{fig:SPDE_3d_dilation}
\end{subfigure}
\caption{Recovery of quadratic statistics for the SPDE \eqref{eq:example_Magnus_Linblad}.
Parameters are $\varepsilon =0.1$, $\beta=0.5$, $\sigma_1=0.5$, $\sigma_2=0.3$, and $T=1.0$.
The evaluation site is $p_* = 5\times 10^{-6}$.
The recovered second moment closely matches the reference solution over the simulated time interval.}
\label{fig:Linblad_recover}
\end{figure}

\begin{figure}
    \centering
    \includegraphics[width=0.5\linewidth]{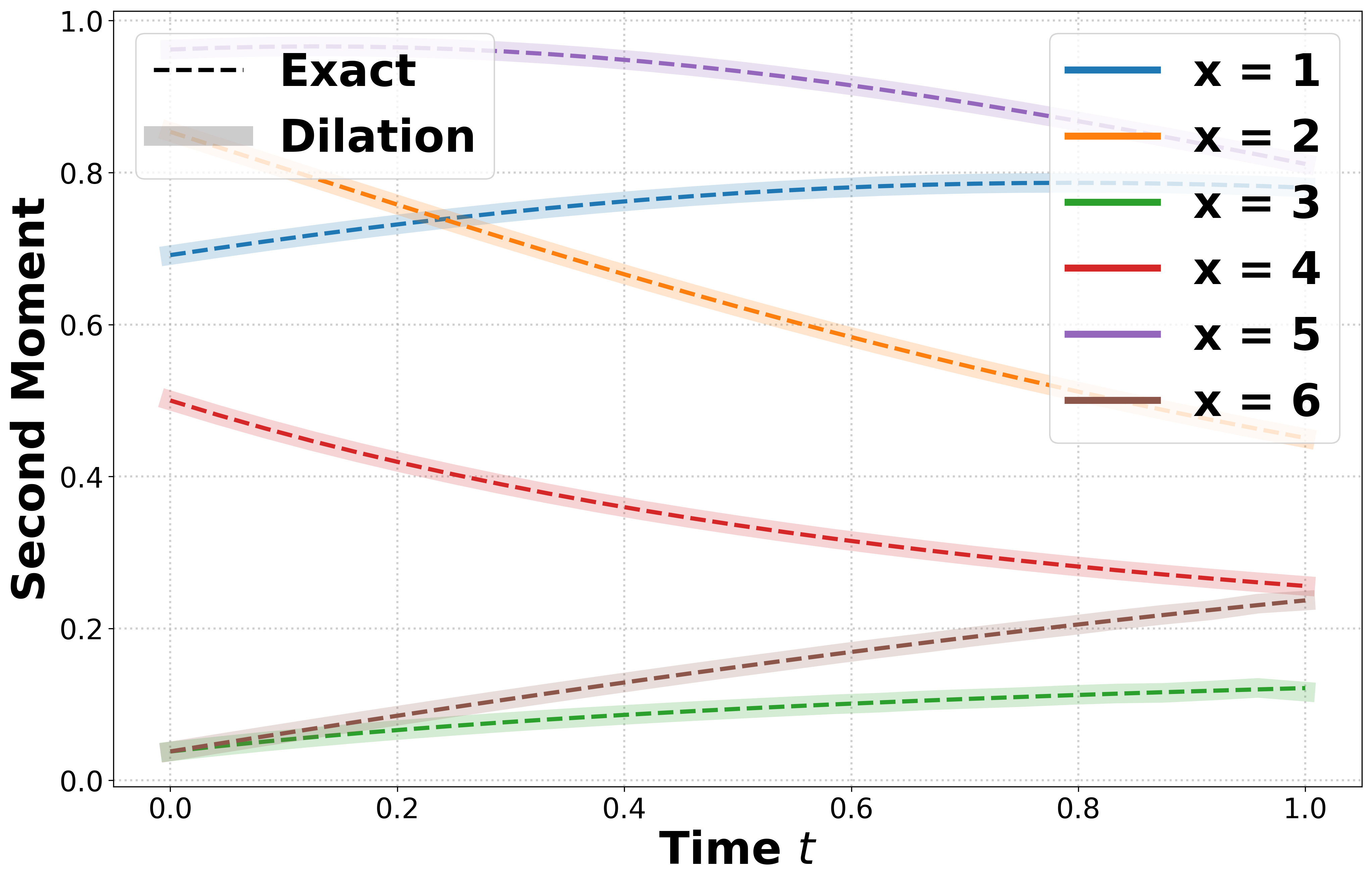}
    \caption{Pointwise comparison of the second moment obtained from the second-moment equation and from the dilation-based recovery.
    Here $p_* = 5\times 10^{-6}$. The agreement improves as $p_*$ decreases, consistent with the localization effect of the projection.}
    \label{fig:compare_second_moment}
\end{figure}

The associated second-moment equation for \cref{eq:example_ito} is
\begin{equation} \label{eq:spde-second-moment}
    \frac{d\Sigma_t}{dt} = A\Sigma_t + \Sigma_t A^\dag + B_1\Sigma_t B_1^\dag + B_2 \Sigma_t B_2^\dag,
    \qquad \Sigma_t\coloneqq\mathbb{E}[X_t X_t^\dag].
\end{equation}
We use \cref{eq:spde-second-moment} as a deterministic reference, and then recover the same quadratic statistics \eqref{eq:Sigma-recover-ideal} using the dilation-based Lindblad simulation described in \cref{lem:Linblad dilated moment}.
\cref{fig:Linblad_recover,fig:compare_second_moment} show that the second moment of the SPDE can be accurately recovered by solving $\rho_t$ in \eqref{eq:lindblad}, as expected.

\section{Summary and discussions}

This work presents an exact mapping from general linear stochastic differential equations (SDEs) to stochastic Schr\"odinger equations (SSEs) through a finite-dimensional dilation framework. A key structural feature is that the dilation can be realized by a nearest-neighbor (tight-binding) hopping operator on the ancilla register, so the resulting circuits are ancilla-efficient and hardware-friendly: the nontrivial ancilla dynamics reduce to local couplings (plus simple boundary terms), and the overall implementation admits a streamlined gate-level construction. 

More broadly, it provides a coherent route for recasting classical stochastic dynamics into quantum-native primitives. By representing the same underlying SDE either as (i) a deterministic Lindblad evolution governing ensemble moments, or as (ii) a measurement-driven quantum-trajectory process generating pathwise realizations, we obtain complementary algorithmic building blocks for stochastic simulation, filtering, data assimilation, forecasting, and sampling. In regimes where these tasks are bottlenecked by repeated propagation of trajectories or by the evolution of high-dimensional moments, quantum implementations can accelerate the dominant inner loops by
enabling long-time propagation through structured dilations and by permitting direct estimation of application-specific observables from the prepared quantum state.

A notable application is the simulation of open quantum systems in the \emph{non-Markovian} regime. Several established trajectory-based methodologies represent non-Markovian effects by embedding the
dynamics into an extended (often higher-dimensional) stochastic model, including hierarchical constructions and Markovian embeddings \cite{suess_hierarchy_2014,li2021markovian,li2023succinct}. Our dilation framework interfaces naturally with these embeddings: once the dynamics are expressed in an extended linear SDE form, the corresponding second-moment evolution is deterministically
captured by a Lindblad equation on the dilated space. This avoids the explicit generation of individual stochastic realizations when only ensemble-level quantities are required, while still retaining the
ability to recover trajectory-level information via the SSE route when needed.

The present paper focuses on Brownian-driven dynamics. Extending the framework to SDEs driven by jump processes, or more generally by L\'evy noise, is a natural next step and will be pursued in
future work. Finally, we note that the trajectory-generation capability (Algorithm II) is particularly relevant to the sampling bottleneck in modern generative models. By mapping the reverse-time SDEs of diffusion models to a dilated quantum evolution, our framework provides a rigorous pathway to accelerate the sampling of high-dimensional distributions beyond the capabilities of classical solvers. To translate this framework into a practical quantum advantage for applications such as diffusion-model sampling, several additional ingredients remain to be developed. These include efficient representations of the problem-specific drift and noise operators, optimized implementations of the resulting Lindblad and trajectory simulators, and a careful end-to-end complexity analysis comparing the full quantum workflow against state-of-the-art classical solvers. We view the current work as providing the mathematical and algorithmic foundation upon which such application-specific quantum accelerations can be built.

\paragraph*{Acknowledgement.}   This research is supported by the NSF Grant DMS-2411120. 

\paragraph*{Author contribution statements.}
All authors contributed equally to this work. The authors acknowledge the use of large language models to assist with the implementation of the numerical experiments and to improve the clarity and grammar of the manuscript.

\bibliographystyle{quantum}
\bibliography{Lindblad,QSDEreferences}

\appendix
\section{Proof of the exact dilation }\label{appendix-A}
\begin{proof}
We begin with the integral form of the SDE for $X_t$:
\begin{equation}
    X_{t} = X_{0} + \int_{0}^{t} A(s)X_{s}ds + \sum_{j=1}^{J}\int_{0}^{t} B_{j}(s)X_{s}dW_{s}^{j}.
\end{equation}
Introducing the notation $B_{0}(t) = A(t)$, $dZ_{t}^{0} = dt$, and $dZ_{t}^{j} = dW_{t}^{j}$ for $j=1, \dots, J$, we rewrite this compactly as:
\begin{equation}
    X_{t} = X_{0} + \sum_{j=0}^{J}\int_{0}^{t} B_{j}(s)X_{s}dZ_{s}^{j}.
\end{equation}
Iterating this integral equation yields the Dyson series expansion:
\begin{equation}\label{eq:Dyson-SDE}
    X_{t} = \sum_{k=0}^{\infty} \left( \sum_{\alpha \in \{0,\dots,J\}^{k}} \int_{0}^{t} \dots \int_{0}^{s_2} B_{\alpha_{k}}(s_k) \dots B_{\alpha_{1}}(s_1) X_{0} \, dZ_{s_1}^{\alpha_1} \dots dZ_{s_k}^{\alpha_k} \right).
\end{equation}
Under the boundedness assumptions on $A,B_j$, this series converges in $L^2(\Omega)$ uniformly on compact time intervals \cite{kloeden}. 

Similarly, the dilated state $\ket{\psi_t}$ evolves according to \cref{eq:dilated-SSE} starting from $\ket{\psi_0} = \ket{r} \otimes X_0$. Its Dyson expansion is:
\begin{equation}\label{eq:Dyson-SSE}
    \ket{\psi_{t}} = \sum_{k=0}^{\infty} \left( \sum_{\alpha \in \{0,\dots,J\}^{k}} \int_{0}^{t} \dots \int_{0}^{s_{2}} V_{\alpha_{k}}(s_k) \dots V_{\alpha_{1}}(s_1) (\ket{r} \otimes X_0) \, dZ_{s_1}^{\alpha_1} \dots dZ_{s_k}^{\alpha_k} \right).
\end{equation}
We now apply the operator $(\bra{l} \otimes I_{\mathcal{A}})$ to the series \cref{eq:Dyson-SSE}. By linearity, it enters the sum and integrals. To evaluate the term-wise action, we separate the vector $X_0$ from the ancilla $\ket{r}$ using the identity $\ket{r} \otimes X_0 = (\ket{r} \otimes I) X_0$:
\begin{equation}
    (\bra{l} \otimes I) \left( V_{\alpha_k} \cdots V_{\alpha_1} \right) (\ket{r} \otimes X_0) 
    = (\bra{l} \otimes I) \left( V_{\alpha_k} \cdots V_{\alpha_1} \right) (\ket{r} \otimes I) X_0.
\end{equation}
Now we use the properties of the moment-matching triple. The sandwiching of the dilated operators yields the original operators:
\begin{equation}
    (\bra{l} \otimes I) (I_{\mathcal{A}} \otimes B_{j}(t)) (\ket{r} \otimes I) = \bra{l} I_\mathcal{A} \ket{r} \otimes B_{j}(t) = B_j(t),
\end{equation}
\begin{equation}
    (\bra{l} \otimes I) V_0(t) (\ket{r} \otimes I) = \bra{l} (F-I_{\mathcal{A}}) \ket{r} \otimes K + \braket{l}{r} \otimes A = A(t).
\end{equation}
Applying this recursively to the product sequence, we obtain,
\begin{equation}
     \left[ (\bra{l} \otimes I) V_{\alpha_k} \cdots V_{\alpha_1} (\ket{r} \otimes I) \right] X_0 
    = \left( B_{\alpha_k} \cdots B_{\alpha_1} \right) X_0.
\end{equation}
Substituting this back into the expansion \cref{eq:Dyson-SSE}, we see that the projection of the quantum state series is exactly \cref{eq:Dyson-SDE}.

\end{proof}

\section{Proof of the light cone property} \label{appendix-B}
\begin{proof}
We proceed in four steps: (1) defining the error dynamics as a driven SSE, (2) expanding the solution using a Dyson series, (3) identifying the non-vanishing terms based on grid locality, and (4) estimating the magnitude of the stochastic integrals.

 A crucial property of any SSE \eqref{eq:dilated-SSE} is that it preserves the norm of the state \textit{on average}. If $\mathcal E(t,s)$ is the propagator for the homogeneous part, then for any state $\phi$:
\begin{equation}\label{eq:norm-preservation}
    \mathbb{E}[ \| \mathcal E(t,s) \phi \|^2 ] = \| \phi \|^2.
\end{equation}
This "mean-square unitarity" simplifies our analysis significantly, as we do not need to worry about the stability of the background evolution.

\medskip

To isolate spatial propagation, we split the homogeneous drift into a ``local'' part and the
nearest-neighbor hopping term,
\[
\widetilde B_0(t)=\widetilde B_0^{(0)}(t)+ V (t),
\qquad
 V (t):=\theta F_h\otimes K(t),
\]
where $\widetilde B_0^{(0)}(t)$ is diagonal in the ancilla basis (and contains the local drift
together with the It\^{o} correction), and $ V (t)$ is the only term that transports
amplitude along the tight-binding chain.

Let $\mathcal{E}_0(t,s)$ denote the propagator of the \emph{local} homogeneous SDE obtained by setting
$ V \equiv 0$ (i.e., keeping $\widetilde B_0^{(0)}$ and all noise terms $\widetilde B_j$).
Then the full propagator $\mathcal{E}(t,s)$, using the variation of constants for SDEs \cite[Theorem 3.1]{mao2007stochastic} repeatedly, admits a Dyson--Duhamel expansion in $ V $, a sum of nested time-ordered integrals
\begin{equation}\label{eq:chi-dyson}
    \ket{\chi_T} \;=\; \sum_{k=0}^{\infty} \ket{\chi_T^{(k)}},
\end{equation}
with the $k$-hop contribution given by the \emph{(k+1)-simplex} integral
\begin{equation}\label{eq:chi-khop}
{
    \ket{\chi_T}^{(k)}
    \;=\;
    \int_{0<s<t_1<\cdots<t_k<T}
    \mathcal{E}_0(T,t_k)\, V (t_k)\,
    \mathcal{E}_0(t_k,t_{k-1})\,\cdots\,
     V (t_1)\,\mathcal{E}_0(t_1,s)\,\ket{S_s}\;
    ds\,dt_1\cdots dt_k.
}
\end{equation}
Equivalently, one may write the same expression as an iterated integral:
\begin{align}
    \ket{\chi_T}^{(k)}
    &=
    \int_{0}^{T}\!\!\left(
      \int_{s}^{T}\!\!\cdots\!\!\int_{t_{k-1}}^{T}
      \mathcal{E}_0(T,t_k) V (t_k)\mathcal{E}_0(t_k,t_{k-1})\cdots
       V (t_1)\mathcal{E}_0(t_1,s)\,dt_k\cdots dt_1
    \right)\ket{S_s}\,ds. \label{eq:chi-khop-iterated}
\end{align}
In \cref{eq:chi-khop}--\cref{eq:chi-khop-iterated}, $ \mathcal{E}_0$ is the propagator of the ``local''
dynamics (diagonal in the ancilla basis), which includes $\widetilde B_0^{(0)}(t)$ and all noise
terms $\widetilde B_j(t)\,dW_t^j$.

We are interested in the projection $(\bra{j_*} \otimes I) \ket{\chi_T}$.
On our nearest-neighbor grid, the hopping operator $\mathcal{V}$ can move an excitation by at most one site. The source $\ket{S_s}$ starts at site $M$. To reach site $j_*$, we must apply $\mathcal{V}$ at least $m = M - j_*$ times.
Therefore, all terms in the Dyson series with order $k < m$ vanish identically. We only need to sum terms with $k \ge m$.

For $k\ge m=M-j_\ast$, recall the simplex representation \eqref{eq:chi-khop}:
\[
\ket{\chi_T}^{(k)}
=
\int_{0<s<t_1<\cdots<t_k<T}
 \mathcal{E}_0(T,t_k) V (t_k) \mathcal{E}_0(t_k,t_{k-1})\cdots
 V (t_1) \mathcal{E}_0(t_1,s)\ket{S_s}\,
ds\,dt_1\cdots dt_k.
\]
Fix $(s,t_1,\dots,t_k)$ and define the random vector
\[
\ket{Y(s,t_1,\dots,t_k)}
:=
 \mathcal{E}_0(T,t_k) V (t_k) \mathcal{E}_0(t_k,t_{k-1})\cdots
 V (t_1) \mathcal{E}_0(t_1,s)\ket{S_s}.
\]
We bound $\mathbb E\left[ \| \ket{\chi_T}^{(k)}\|^2 \right]$ by applying (i) Cauchy-Schwarz in the
time variables and (ii) the mean-square isometry of $ \mathcal{E}_0(\cdot,\cdot)$ \emph{stepwise} using
conditional expectations.

\medskip
First,  since $ \mathcal{E}_0(t,u)$ is the propagator of the homogeneous ``local'' SSE (with drift in the SSE
form and noise terms $\widetilde B_j$), it is mean-square norm preserving in the following
conditional sense: for $u\le t$ and any $\mathcal F_u$-measurable random vector $\zeta$,
\begin{equation}\label{eq:cond-isometry}
    \mathbb E\!\left[\| \mathcal{E}_0(t,u)\zeta\|^2\,\middle|\,\mathcal F_u\right]
    =\|\zeta\|^2
    \qquad\text{a.s.}
\end{equation}
(Equivalently, $\mathbb E\| \mathcal{E}_0(t,u)\zeta\|^2=\mathbb E\|\zeta\|^2$ by taking expectations.)
Identity~\eqref{eq:cond-isometry} follows from It\^o's formula applied to
$\|\eta_t\|^2$ for the homogeneous local SSE and the fact that the Brownian increments on
$(u,t]$ are independent of $\mathcal F_u$.

\medskip
Next, we simplify the nested integral using this isometry. Define
\[
\zeta_k:= V (t_k) \mathcal{E}_0(t_k,t_{k-1})\cdots  V (t_1) \mathcal{E}_0(t_1,s)\ket{S_s},
\]
which is $\mathcal F_{t_k}$-measurable. Applying ~\eqref{eq:cond-isometry} with $(t,u)=(T,t_k)$ yields
\[
\mathbb E\!\left[\|\ket{Y(s,t_1,\dots,t_k)}\|^2 \,\middle|\,\mathcal F_{t_k}\right]
=
\mathbb E\!\left[\| \mathcal{E}_0(T,t_k)\zeta_k\|^2 \,\middle|\,\mathcal F_{t_k}\right]
=\|\zeta_k\|^2.
\]
Taking expectations gives
\[
\mathbb E\|\ket{Y(s,t_1,\dots,t_k)}\|^2=\mathbb E\|\zeta_k\|^2.
\]
Now define $\zeta_{k-1}:= V (t_{k-1}) \mathcal{E}_0(t_{k-1},t_{k-2})\cdots V (t_1) \mathcal{E}_0(t_1,s)\ket{S_s}$, which is $\mathcal F_{t_{k-1}}$-measurable. Since $\zeta_k= V (t_k) \mathcal{E}_0(t_k,t_{k-1})\zeta_{k-1}$,
taking conditional expectation with respect to $\mathcal F_{t_{k-1}}$ gives us
\begin{align*}
\mathbb E\|\zeta_k\|^2
&=
\mathbb E\!\left[\,
\mathbb E\!\left(\| V (t_k) \mathcal{E}_0(t_k,t_{k-1})\zeta_{k-1}\|^2 \,\middle|\,\mathcal F_{t_{k-1}}\right)
\right]\\
&\le
\| V (t_k)\|^2\,
\mathbb E\!\left[\,
\mathbb E\!\left(\| \mathcal{E}_0(t_k,t_{k-1})\zeta_{k-1}\|^2 \,\middle|\,\mathcal F_{t_{k-1}}\right)
\right]\\
&=
\| V (t_k)\|^2\,\mathbb E\|\zeta_{k-1}\|^2,
\end{align*}
where in the last line we used \eqref{eq:cond-isometry}. Iterating this argument yields the
``propagator peeling'' bound
\begin{equation}\label{eq:peel}
\mathbb E\|\ket{Y(s,t_1,\dots,t_k)}\|^2
\le
\left(\prod_{\ell=1}^k \| V (t_\ell)\|^2\right)\,
\mathbb E\| \mathcal{E}_0(t_1,s)\ket{S_s}\|^2.
\end{equation}
Finally, applying \eqref{eq:cond-isometry} once more with $(t,u)=(t_1,s)$ gives
$\mathbb E\| \mathcal{E}_0(t_1,s)\ket{S_s}\|^2=\mathbb E\|\ket{S_s}\|^2$. Therefore,
\begin{equation}\label{eq:Y-bound}
\mathbb E\|\ket{Y(s,t_1,\dots,t_k)}\|^2
\le
\left(\prod_{\ell=1}^k \| V (t_\ell)\|^2\right)\,
\mathbb E\|\ket{S_s}\|^2.
\end{equation}

\medskip
Finally, 
using $\| V (t)\|\le \theta\|F_h\|\,\|K(t)\|\le \theta\,2 C_{grid}\,K_{\max}$, 
where \[C_{grid} := \sup_{h\geq 1} \frac{\sqrt{1+e^h}}{4 \sinh(h /2)} <1. \] 
With direct computation, we can show that 
$\mathbb E\|\ket{S_s}\|^2\le 2 K_{\max}^2 X(T)$ assuming 
\begin{equation}\label{eq:XT-def}
    X(T)\;:=\;\sup_{0\le t\le T}\mathbb{E}\,\|X_t\|^2,
\end{equation}
can come from the stability estimates \cref{prop:stability}.

Inequality~\eqref{eq:Y-bound} implies
\[
\mathbb E\|\ket{Y(s,t_1,\dots,t_k)}\|^2
\le
\bigl(\theta C_{grid}K_{\max}\bigr)^{2k} K_{\max}^2  X(T).
\]
Applying Cauchy--Schwarz in the time variables gives
\begin{align*}
    \mathbb E\|\ket{\chi_T}^{(k)}\|^2
&=
\mathbb E\left\|\int_{\Delta_{k+1}(T)} \ket{Y(s,t_1,\dots,t_k)}\,d(s,t_1,\dots,t_k)\right\|^2 \\
&\le
|\Delta_{k+1}(T)|\int_{\Delta_{k+1}(T)} \mathbb E\|\ket{Y}\|^2\,d(s,t_1,\dots,t_k),
\end{align*}
where $\Delta_{k+1}(T)=\{0<s<t_1<\cdots<t_k<T\}$ and $|\Delta_{k+1}(T)|=T^{k+1}/(k+1)!$.
Hence
\begin{equation}\label{eq:chi-k-bound}
\mathbb E\|\ket{\chi_T}^{(k)}\|^2
\le
 K_{\max}^2  X(T) \left(\theta C_{grid}K_{\max}\right)^{2k}\left(\frac{T^{k+1}}{(k+1)!}\right)^2.
\end{equation}

Using Stirling's approximation $(k+1)!\ge \bigl(\tfrac{k+1}{e}\bigr)^{k+1}$, we obtain the
geometric decay for $k\ge m$:
\[
\mathbb E\|\ket{\chi_T}^{(k)}\|^2
\le
K_{\max}^2  X(T)
\left(\frac{e\,\theta C_{grid}K_{\max}T}{k+1}\right)^{2k}
\cdot \left(\frac{T}{k+1}\right)^2
\;\lesssim\;
K_{\max}^2  X(T)\varrho^{2k},
\]
where $
\varrho:=\frac{e\,\theta C_{grid}K_{\max}T}{m}.$

The different hopping orders $\ket{\chi_T^{(k)}}$ are not necessarily
orthogonal in $L^2(\Omega)$, so we do not sum the squared norms directly.
Instead, we apply Minkowski's inequality before squaring:
\[
\left(\mathbb E\|\ket{\chi_T}\|^2\right)^{1/2}
=
\left(\mathbb E\left\|\sum_{k=m}^{\infty}\ket{\chi_T^{(k)}}\right\|^2\right)^{1/2}
\le
\sum_{k=m}^{\infty}
\left(\mathbb E\|\ket{\chi_T^{(k)}}\|^2\right)^{1/2}.
\]
Using the per-order estimate above gives
\[
\left(\mathbb E\|\ket{\chi_T}\|^2\right)^{1/2}
\lesssim
K_{\max}\sqrt{X(T)}
\sum_{k=m}^{\infty}\varrho^k
=
K_{\max}\sqrt{X(T)}
\frac{\varrho^m}{1-\varrho}.
\]
Therefore,
\[
\mathbb E\|\ket{\chi_T}\|^2
\le
\mathcal C
\frac{\varrho^{2m}}{(1-\varrho)^2},
\]
for a constant $\mathcal C$ depending on $K_{\max}$, $X(T)$, $T$, and grid
boundary constants. This proves the claimed light-cone decay without assuming
orthogonality between different hopping orders.

\end{proof}

\section{Proof of the light-cone property for the covariance} \label{sec: lcc}
\begin{proof}
Let $\ket{\phi_t} = \ket{r_h} \otimes X_t$ be the ideal dilated state (solution to the SDE embedded in the dilated space).
Let $\ket{\psi_t}$ be the actual dilated state evolving under the dilated SSE. The error state is defined as $\ket{\chi_t} = \ket{\psi_t} - \ket{\phi_t}$. We aim to bound the trace distance between the actual projected density matrix, $\rho_T^{(j_*)}$, and the ideal covariance block, $|\gamma|^2 \Sigma_T$.

The full density matrix is defined as $\rho_T = \E{\ket{\psi_T}\bra{\psi_T}}$. Substituting $\ket{\psi_T} = \ket{\phi_T} + \ket{\chi_T}$, we expand the outer product:
\begin{align}
    \rho_T &= \E{(\ket{\phi_T} + \ket{\chi_T})(\bra{\phi_T} + \bra{\chi_T})} \nonumber \\
    &= \underbrace{\E{\ket{\phi_T}\bra{\phi_T}}}_{\text{Ideal Term}} 
     + \underbrace{\E{\ket{\phi_T}\bra{\chi_T}} + \E{\ket{\chi_T}\bra{\phi_T}}}_{\text{Cross Terms}} 
     + \underbrace{\E{\ket{\chi_T}\bra{\chi_T}}}_{\text{Quadratic Error}}
\end{align}
Let $\mathcal{P}_{j_*} \coloneqq \bra{j_*} \otimes I$ be the projection operator onto the ancilla site $j_*$. We analyze the projection of each term separately.

We first look at the ideal term. Recalling that $\ket{\phi_T} = \ket{r_h} \otimes X_T$ and $\gamma \coloneqq \braket{j_*}{r_h}$. We have
\begin{equation}
    \mathcal{P}_{j_*} \E{\ket{\phi_T}\bra{\phi_T}} \mathcal{P}_{j_*}^\dagger 
    = \E{(\braket{j_* }{r_h} X_T)(\braket{j_* }{r_h} X_T^\dagger)} \nonumber = |\gamma|^2 \E{X_T X_T^\dagger} \nonumber = |\gamma|^2 \Sigma_T.
\end{equation}

Now we evaluate the bound for the error terms. Let the projected state vectors be denoted as $\ket{\phi}_{j_*} \coloneqq \mathcal{P}_{j_*} \ket{\phi_T}$ and $\ket{\chi}_{j_*} \coloneqq \mathcal{P}_{j_*} \ket{\chi_T}$. The error in the projected density matrix is:
\begin{equation}
    \rho_T^{(j_\ast)}-|\gamma|^2\Sigma_T = \E{\ket{\phi}_{j_*}\bra{\chi}_{j_*}} + \E{\ket{\chi}_{j_*}\bra{\phi}_{j_*}} + \E{\ket{\chi}_{j_*}\bra{\chi}_{j_*}}.
\end{equation}
We apply the trace norm $\norm{\cdot}_1$ and the triangle inequality:
\begin{equation}
    \norm{\rho_T^{(j_\ast)}-|\gamma|^2\Sigma_T}_1 \le 2 \norm{\E{\ket{\phi}_{j_*}\bra{\chi}_{j_*}}}_1 + \norm{\E{\ket{\chi}_{j_*}\bra{\chi}_{j_*}}}_1 .
\end{equation}
We obtain the following inequality by using Jensen's inequality and the property that $\norm{\ket{u}\bra{v}}_1 = \norm{u}\norm{v}$,
\begin{equation}
    \norm{\E{\ket{\phi}_{j_*}\bra{\chi}_{j_*}}}_1 \le \E{\norm{\ket{\phi}_{j_*}\bra{\chi}_{j_*}}_1} = \E\left[{\norm{\ket{\phi}_{j_*}} \cdot \norm{\ket{\chi}_{j_*}}}\right].
\end{equation}
We further apply the Cauchy-Schwarz inequality for expectations,
\begin{equation}
    \E\left[{\norm{\ket{\phi}_{j_*}} \cdot \norm{\ket{\chi}_{j_*}}}\right] \le \sqrt{\E{\norm{\ket{\phi}_{j_*}}}^2} \cdot \sqrt{\E{\norm{\ket{\chi}_{j_*}}}^2}.
\end{equation}
Note that $\norm{\ket{\phi}_{j_*}} = |\gamma|\norm{X_T}$, combining with the result from \cref{thm:light-cone}, we get
\begin{align}
    \sqrt{\E{\norm{\ket{\psi}_{j_*}}^2}} &= |\gamma| \sqrt{\E{\norm{X_T}^2}} = |\gamma| \sqrt{\tr(\Sigma_T)}, \\
    \sqrt{\E{\norm{\ket{\chi}_{j_*}}^2}} &\le \sqrt{\mathcal{C}\frac{\varrho^{2m}}{(1-\varrho)^2}}.
\end{align}
Thus, the contribution of the cross terms is bounded by
\begin{equation} \label{eq:bound_cross_err}
    2 \norm{\E{\ket{\phi}_{j_*}\bra{\chi}_{j_*}}}_1 \le 2 \cdot |\gamma| \sqrt{\tr(\Sigma_T)} \sqrt{\mathcal{C}\frac{\varrho^{2m}}{(1-\varrho)^2}}.
\end{equation}

Since $\ket{\chi}_{j_*}\bra{\chi}_{j_*}$ is positive semi-definite, its trace norm equals its trace, we obtain
\begin{equation} \label{eq:bound_quad err}
    \norm{\E{\ket{\chi}_{j_*}\bra{\chi}_{j_*}}}_1 = \tr\left(\E{\ket{\chi}_{j_*}\bra{\chi}_{j_*}}\right) = \E\left[ {\tr(\ket{\chi}_{j_*}\bra{\chi}_{j_*})} \right] = \E\left[{\norm{\ket{\chi}_{j_*}}^2}\right] \le \mathcal{C}\frac{\varrho^{2m}}{(1-\varrho)^2}.
\end{equation}
The proof is completed by summing up the bounds in \cref{eq:bound_cross_err} and \cref{eq:bound_quad err}.
\end{proof}

\section{Proof of the weak order 2 dilation}\label{appendix-order2}
\begin{proof}
Write \(K:=\Omega_n\), \(U:=U_n\), and abbreviate \(G_\alpha:=G_{\alpha,n}\).
A direct computation using orthogonality of \(\ket{10},\ket{01},\ket{11}\) yields, for any
\(\ket{\psi}\),
\begin{align}
K(\ket{00}\otimes\ket{\psi})
&=\sum_{\alpha\in\{10,01,11\}}\ket{\alpha}\otimes(G_\alpha\ket{\psi}),\label{eq:K1-3ch-proof}\\
K^2(\ket{00}\otimes\ket{\psi})
&=-\ket{00}\otimes\big(R^2\ket{\psi}\big),
\qquad
R^2:=\sum_{\alpha\in\{10,01,11\}}G_\alpha^\dagger G_\alpha.\label{eq:K2-3ch-proof}
\end{align}
Iterating gives the standard even/odd pattern
\(K^{2q}(\ket{00}\otimes\ket{\psi})=(-1)^q\ket{00}\otimes(R^{2q}\ket{\psi})\) and
\(K^{2q+1}(\ket{00}\otimes\ket{\psi})=(-1)^q\sum_\alpha \ket{\alpha}\otimes(G_\alpha R^{2q}\ket{\psi})\).
Therefore the first column of \(U=e^K\) admits the truncated series
\begin{equation}\label{eq:U-column-trunc-proof}
U(\ket{00}\otimes\ket{\psi})
=
\ket{00}\otimes\Omega\,\ket{\psi}
+\sum_{\alpha\in\{10,01,11\}}\ket{\alpha}\otimes\Big(G_\alpha\Lambda\,\ket{\psi}\Big)
+\mathcal R\,\ket{\psi},
\end{equation}
where
\(
\Omega:=I-\tfrac12 R^2+\tfrac1{24}R^4
\),
\(
\Lambda:=I-\tfrac16 R^2
\),
and \(\|\mathcal R\|=\co(\Delta t^{5/2})\) because \(\|G_1\|=\co(\Delta t^{1/2})\), \(\|G_2\|=\co(\Delta t)\),
\(\|G_3\|=\co(\Delta t^{3/2})\).

Now fix presampled \((\xi_1,\xi_2)\) and set \(\xi_3=\xi_1^2-1\), \(\ket{m}:=\ket{m(\xi_1,\xi_2)}\).
By construction,
\begin{equation}\label{eq:ratio-xi123-proof}
\frac{\braket{m}{10}}{\braket{m}{00}}=\xi_1,\qquad
\frac{\braket{m}{01}}{\braket{m}{00}}=\frac{\xi_3}{\sqrt2},\qquad
\frac{\braket{m}{11}}{\braket{m}{00}}=\xi_2.
\end{equation}
Project \cref{eq:U-column-trunc-proof} with \(\bra{m}\) and divide by \(\braket{m}{00}\) to obtain
\begin{equation}\label{eq:Keff-ratio-form-proof}
\widetilde K
=
\Omega
+\xi_1\,G_1\Lambda
+\frac{\xi_3}{\sqrt2}\,G_2\Lambda
+\xi_2\,G_3\Lambda
\;+\;\co_{\rm w}(\Delta t^3).
\end{equation}

It remains to expand each term to the weak-$2$ relevant orders. First,
\(R^2=G_1^\dagger G_1+\co(\Delta t^2)=\Delta t\,V_{\rm mid}^\dagger V_{\rm mid}+\co(\Delta t^2)\),
hence
\(\Omega=I-\tfrac{\Delta t}{2}V_{\rm mid}^\dagger V_{\rm mid}+\co(\Delta t^2)=I+\Delta t\,A_{\rm mid}+\co(\Delta t^2)\),
and \(\Lambda=I-\tfrac{\Delta t}{6}V_{\rm mid}^\dagger V_{\rm mid}+\co(\Delta t^2)\).
Using the choice \eqref{eq:G123-choice-rev}, we then obtain
\begin{align*}
\xi_1\,G_1\Lambda
&=
\xi_1\Big(\sqrt{\Delta t}\,V_{\rm mid}+\Delta t^{3/2}\big(B_{\rm mid}+\tfrac16 V_{\rm mid}V_{\rm mid}^\dagger V_{\rm mid}\big)\Big)
\Big(I-\tfrac{\Delta t}{6}V_{\rm mid}^\dagger V_{\rm mid}\Big)
+\co(\Delta t^{5/2})\\
&=
\sqrt{\Delta t}\,\xi_1\,V_{\rm mid}
+\Delta t^{3/2}\,\xi_1\,B_{\rm mid}
+\co(\Delta t^{5/2}),
\end{align*}
where the \(\tfrac16 V_{\rm mid}V_{\rm mid}^\dagger V_{\rm mid}\) term cancels the \(-\tfrac16 V_{\rm mid}V_{\rm mid}^\dagger V_{\rm mid}\)
contribution induced by \(\Lambda\).
Similarly,
\[
\frac{\xi_3}{\sqrt2}\,G_2\Lambda
=
\frac{\xi_3}{\sqrt2}\cdot \frac{\Delta t}{\sqrt2}V_{\rm mid}^2\cdot\big(I+\co(\Delta t)\big)
=
\frac{\Delta t}{2}\,\xi_3\,V_{\rm mid}^2
+\co_{\rm w}(\Delta t^3),
\qquad
\xi_2\,G_3\Lambda
=
\Delta t^{3/2}\,\xi_2\,C_{\rm mid}
+\co_{\rm w}(\Delta t^3).
\]
Finally, the deterministic \(\co(\Delta t^2)\) contribution produced by \(\Omega\) (coming from the
\(-\tfrac12 R^2\) and \(\tfrac1{24}R^4\) terms, and from the \(\co(\Delta t^2)\) part of \(R^2\))
matches \(\tfrac{\Delta t^2}{2}A_{\rm mid}^2\) up to an \(\co_{\rm w}(\Delta t^3)\) weak remainder; this is
precisely the same weak-$2$-preserving midpoint simplification used in passing from
\cref{eq:weak2-ito-taylor-tn-clean} to \cref{eq:weak2-ito-taylor-mid-clean}.
Collecting all terms gives \cref{eq:weak2-expansion-one-dilation}.
\end{proof}

\end{document}